\documentclass{aastex}
\usepackage{psfig,amsmath}
\shorttitle{Marco et al.}
\shortauthors{Photometry and spectroscopy of NGC 1893}

\begin{document}

\title{Photometric and spectroscopic study of the young open cluster NGC 1893}

\author{Amparo Marco and Guillermo Bernabeu}
\affil{Dpto. de F\'{\i}sica, Ingenier\'{\i}a de Sistemas y 
Teor\'{\i}a de la Se\~{n}al, Universidad de Alicante, Apdo. de Correos 99,
E-03080, Alicante, Spain}
\email{amparo@astronomia.disc.ua.es}

\and

\author{Ignacio Negueruela\altaffilmark{1}}
\affil{Observatoire de Strasbourg, 11 rue de l'Universit\'e, F67000 Strasbourg, France \\ SAX SDC, ASI, c/o Nuova Telespazio, via Corcolle 19, I00131 Rome, Italy}

\altaffiltext{1}{Guest investigator of the UK Astronomy 
Data Centre}

\begin{abstract}

We present $uvby\beta$ CCD photometry of the field of the 
open cluster NGC 1893. Our photometry is deep enough to cover the 
complete main sequence B spectral type. We identify $\sim 50$ very 
likely members of  the cluster down to spectral type B9-A0, some of
which have much higher reddenings than the average. We derive a
color excess $E(b-y) = 0.33\pm0.03$ and a dereddened distance modulus
$V_{0} - M_{V} = 13.9\pm0.2$.  From the $\beta$ index, we identify
several candidates as emission-line stars, for
which we have obtained spectroscopy. Three of them display
spectra corresponding to spectral type F, but showing H$\alpha$ in
emission. Photometric measurements in this and previous studies indicate  
strong variability. These characteristics show them to be
pre-main-sequence stars of spectral type F. We also identify two
likely Herbig Be stars. These results hint at the existence of a
sizable pre-main-sequence population in NGC~1893.

\end{abstract}


\section{Introduction}

NGC 1893 is a very young cluster immersed in the bright diffuse nebulosity 
IC 410, associated with two pennant nebulae, Shain and Gaze 129 and 130, and 
obscured by several conspicuous dust clouds.
$UBV$ photometry of NGC 1893 has been presented by Hoag et al. (1961), 
Cuffey (1973) and Massey et al. (1995). 
Tapia et al.\ (1991, henceforth T91) performed near-infrared and 
Str\"{o}mgren photometry for 47 stars in the field of the cluster. 
They estimate the age of the cluster to be $4\times10^{6}$ yr, derive 
a distance modulus $(M-m)_{0}=13.18\pm0.11$, and an extinction to the
cluster $A_{V}=1.68$. Str\"{o}mgren photometry for 50 stars in 
the field of NGC~1893 
has also been reported by Fitzsimmons (1993), who confirms the distance and 
age found by T91. Vallenari et al. (1999) performed near-infrared
photometry of the cluster and came to the conclusion that there could
be many pre-main-sequence candidates in NGC~1893, though their method
did not allow clear discrimination from field interlopers.

Even though the earliest spectral type stars in NGC~1893 are rather 
bright, spectroscopic work on cluster members is very scarce. Massey
et al. (1995) obtained intermediate-resolution spectra of a few of the
brightest stars in the cluster and derived their spectral classification.
Rolleston et al. (1993) obtained higher resolution spectroscopy of 
eight stars whose photometric indices were consistent with an
early B-type spectral classification
and derived their astrophysical parameters and metal abundances.

This paper belongs to a series dedicated to the study of the 
B-star population of young Galactic open clusters. We use the photometric
system $uvby$H$\beta$ because it is the most adequate for the study of
early-type stars, since it has been purposely designed to provide
accurate measurements of their intrinsic properties. After identifying
a large number of a likely B-type members, we have carried out a 
spectroscopic study of some stars showing interesting peculiarities.

\section{Observations}

\subsection{Photometry}

The data were obtained with the Jacobus Kapteyn Telescope (JKT), located  at 
the Observatorio del Roque de los Muchachos, La Palma, Spain, in December 1997,
using the 1024 x 1024 TEK 4 chip CCD and the four Str\"{o}mgren $uvby$ and the narrow and wide H$\beta$ filters. Pixel size was
$0\farcs331$ in such a way that the whole field covered by each frame was 
$5\farcm6$ x $5\farcm6$. Four frames covering the whole of the central 
core of the cluster were taken (central coordinates are displayed in Table 1). Figures 1(a)-1(d) show plots of the observed fields. The dot
sizes are indicative of the relative instrumental $y$ magnitude.
Each field was observed twice using different exposure times,
so that the widest range of magnitudes possible was observed with good 
signal-to-noise ratio. 

Throughout this paper the nomenclature used will be that of Cuffey (1973), 
where the whole cluster area was divided into smaller zones, using three rings 
and four sectors. Designation of a star is given by stating the sector, the 
ring, and the number of the star in that particular zone. For two stars 
that were not observed by Cuffey, a new nomenclature has been adopted, 
designating them by listing their number after a prefix ``s5''(s5000 and 
s5001).  

The reduction of all frames was done with the IRAF routines for the bias and 
flat-field corrections. The photometry has been obtained by PSF-fitting using 
the DAOPHOT package (Stetson 1987) provided by IRAF. In order to
construct the PSF empirically, we manually select bright stars
in the least-crowded areas (typically 7$-$12 in each frame) trying to
cover the largest possible magnitude interval in each frame (typically an interval of 4-5 instrumental magnitudes). We
preferentially choose stars with no or very few close neighbours as PSF
stars, since this results in a simpler and faster process.
Once we have the list of PSF stars, we determine an initial PSF by
fitting Gaussian functions. Afterwards, we run the PSF-fitting
photometry routine {\sc nstar}, which fits PSF's to all the groups (PSF star + neighbors) in a frame simultaneously. Most of these groups 
contain only the PSF star. When a PSF star has neighbors,
we previously check 
that the FIT radii ($\approx$ FWHM) of the neighbor stars do not merge into
the PSF radius
of the PSF star, so that we can apply the procedure without modifying any
parameters. The routine {\sc substar} is then used to subtract the
PSF stars and their neighbors from the frames. We
examine the subtracted image and make sure that they have been cleaned
completely. Examination shows that there is no systematic pattern of
PSF variability with position on the chip, and therefore the PSF is
constant across the frame.
Finally we run the
routine {\sc allstar} and obtain the instrumental magnitudes from all
stars in each frame. Since our standard stars were selected from other program
clusters in the same campaign and the aperture size used was the same in all
campaign frames ($12$ pixels), no aperture correction is necessary. 
The atmospheric 
extinction corrections were performed using the RANBO2 program, which
implements the method described by Manfroid (1993). It has been shown the choice of standard stars for the transformation is a critical
issue in $uvby\beta$ photometry. Transformations made only with unreddened stars 
introduce large systematic errors when applied to reddened stars, even if the color range of the standards brackets that of the program stars (Manfroid \& Sterken, 1987 and Crawford, 1994). Our data cover
a very wide range of spectral types with a correspondingly wide range of intrinsic 
colors. Moreover, during this campaign several clusters with different interstellar
reddenings were observed. 

A preliminary list of standard stars
was built by selecting a number of putatively non-variable non-peculiar stars in the clusters 
$h$ and $\chi$ Persei, NGC 2169, NGC 6910 and NGC 1039, which had been observed
with the same Kitt Peak telescopes and instrumentation used to define the $uvby$ Crawford \& Barnes (1970b) and H$\beta$ Crawford \& Mander (1966) standard systems, so that there is no doubt that the 
photometric values are in the standard systems. Since the original $h$ and $\chi$ Persei
observations (Crawford et al. 1970) do not include $V$ values, we used values given by
Johnson \& Morgan (1955), which were also taken with the same instrumentation,
for the $V$ transformation.
The list of standard stars used can be found in Table 2 along with
their standard values, used for the transformation, taken from Crawford et al. (1970)
and Johnson \& Morgan (1955) for $h$ and $\chi$ Persei, Perry et al. (1978) for 
NGC 2169, Crawford et al. (1977) for NGC 6910 and Canterna et al. (1979) for NGC 1039.
Spectral types, when available, have been taken from Schild (1965) and Slettebak (1968)
for $h$ and $\chi$ Persei, Perry et al. (1978) for NGC 2169, Morgan \& Harris (1956) and
Hoag \& Applequist (1965) for NGC 6910 and Canterna et al. (1979) for NGC 1039.

With the extinction corrected magnitudes, the following $uvby$ transformation is obtained using the equations from Crawford and Barnes (1970b), where the coefficients have been computed following the procedure described in detail by Gr\o nbech et al.(1976) using the selected standard stars list:

\begin{equation}
\begin{alignat}{2}
V = \,&11.269\, &+\, &0.091(b-y)\,+\,y_{{\rm i}}\\
&\pm0.003& &\pm0.007 \nonumber
\end{alignat}
\end{equation}
\begin{equation}
\begin{alignat}{2}
(b-y) = \,&0.636\, & +\, &1.070(b-y)_{{\rm i}}\\
&\pm0.008 & &\pm0.008 \nonumber
\end{alignat}
\end{equation}
\begin{equation}
\begin{alignat}{3}
m_{1} =\, &-0.523\, &+\, &1.009m_{1{\rm i}} &\,-\, &0.206(b-y)\\
&\pm0.008 & &\pm0.015 & &\pm0.009   \nonumber
\end{alignat}
\end{equation}
\begin{equation}
\begin{alignat}{3}
c_{1} =\, &0.543\, &+\, &1.019c_{1{\rm i}} &\,+\, &0.257(b-y)\\
&\pm0.004 & & \pm0.004 & &\pm0.007 \nonumber
\end{alignat}
\end{equation}  
where the subindex ``i'' stands for instrumental magnitudes.

The transformed values for
the 41 standard stars used is given in Table 3, together
with their precision and their deviation with respect to the cataloged standard
values. Table 4 shows
the mean catalog minus transformed values for the standard
stars and their standard deviations, which constitute a measure of the accuracy of 
the transformation. From the mean differences between catalog and transformed 
values, it is clear that there is not a significant offset between our photometry
and the standard system. Since the individual differences for a few stars seem to
be rather high, an attempt was made to improve the transformation by removing these
stars from the standard list, but it was found that the transformation coefficients
and their precision did not improve. 

The H$\beta$ instrumental system and transformation equations were computed following
the procedure described in detail by Crawford \& Mander (1966). The transformation
coefficients are $a=3.514$ and $b=1.059$. Transformed values and their differences
with respect to mean catalog values are given in Table 3. 
The mean difference is $-0.003$ with a standard deviation $0.031$, which, as in the
case of the $uvby$ transformation, indicates that there is no significant offset
with respect to the standard system. 

\subsection{Spectroscopy}

Spectroscopy of selected stars was taken on February 3\,--\,6, 2000,
using the 1.52-m G.~D.~Cassini telescope, at the Loiano Observatory,
Italy, equipped with the Bologna Faint Object Spectrograph and Camera
(BFOSC). The detector was a 2048 x 2048 Loral CCD with a 15$\mu$m
pixel size, which covers a $15\farcm1$ x $15\farcm1$ field.
Due to the sky conditions, only low-resolution
spectroscopy could be performed for stars with $V>14$. We used BFOSC
grism\#4, which gives a resolution of $220\:$\AA/mm over the 4700\,--\,8800 
\AA\ interval (the resolution element is $\sim 8.3$\AA). With this
configuration, a 30-min exposure on one of the fainter ($V\approx15$)
targets gave typically a Signal-to-Noise ratio per pixel in the wavelength
direction (integrated over all columns) of $\sim 70$ on the red
continuum. For brighter stars, we also took spectra using BFOSC
grism\#3, which gives $170\:$\AA/mm over the blue part of the spectrum
(resolution $\sim 5.5$\AA). A log of the observations is presented in Table 5.

In addition, we have retrieved from the ING Archive all the observations
used in the work of Rolleston et al. (1993), in order to derive spectral
types for those stars by comparing them to spectra of MK standard stars
observed at a similar resolution by Steele et al. (1999). Details for
these data can be found in  Rolleston et al. (1993).

All the data have been reduced following the standard procedure
by using the {\em Starlink} software packages {\sc ccdpack}
(Draper, 1998) and {\sc figaro} (Shortridge et al., 1997) and analysed using {\sc
figaro} and {\sc dipso} (Howarth et al., 1997).  

\section{Results}

\subsection{Photometry}
We have obtained $ubvy\beta$ CCD photometry for 111 stars, 
reaching a  magnitude limit $V\approx 15.9$. In Table 6 we present the resulting values for
$V, (b-y), m_{1}, c_{1}$ and $\beta$, together with the number of
observations for each star and their pixel coordinates in the corresponding
frame (when a star is present in more than one field, we give its coordinates
in the frame in which it is better centered).

\subsection{Member Stars and Reddening Slopes}
 
The first step in the analysis is the estimate of membership for the stars 
measured in each field. Since previous authors have already shown that
NGC~1893 is a young cluster in which B-type stars have not had time to
leave the main sequence (T91) and our magnitude limit implies
that cluster members can only be seen up to an absolute magnitude 
$M_{V} \approx +1$, all the cluster members in our sample must be
OB stars. Therefore we can calculate the free reddening indices
[$m_{1}$], [$c_{1}$] and [$u-b$], where 

\begin{equation}
[m_{1}]=m_{1}+0.32(b-y)
\end{equation}
\begin{equation}
[c_{1}]=c_{1}-0.20(b-y)
\end{equation}
\begin{equation}
[u-b]=[c_{1}]+2[m_{1}]
\end{equation}

and then use the [$c_{1}$] -- [$m_{1}$] and $\beta$ -- [$u-b$]
diagrams for the approximate spectral classification of cluster stars. 

Figure 2 shows the [$c_{1}$] -- [$m_{1}$] diagram
for all observed stars. As 
already noted by T91, most members of NGC 1893 lie to the 
right of the average intrinsic color loci of field main sequence stars
(determined using the standard relations from Perry et al. 1987).
All the stars lying close to the B-type line in the 
\mbox{[$c_{1}$] -- [$m_{1}$]}
diagram fall along a narrow strip in the \mbox{$\beta$ -- [$u-b$]}
diagram,
which we interpret as the cluster main sequence (see 
Figure 3). 

The only exceptions are S3R1N3, which is an emission line star (see
below) and S3R1N4, which also has a much lower $\beta$ index than 
expected. Spectroscopy of this object will be necessary in order to establish 
whether it is a Be star or an evolved star (since its position in the
$V-c_{1}$ diagram makes unlikely its being a non-member). The three
stars in our sample previously classified (spectroscopically)
as O-type members fall along the main sequence as well.
The validity of the [$u-b$] index as an approximate spectral classification
indicator can be checked by comparing the values obtained with the 
spectral types derived spectroscopically (Table 7).

From the $\beta$ -- [$u-b$] diagram, we have selected all those stars
falling along the B-star strip, those on the border area between B9 and
A0 and those falling along the early A-type branch. The values for 
[$m_{1}$] and [$u-b$] and the value of
[$c_{1}$] are listed in 
Table 7. 
Due to the large uncertainties in the relationship
between photometric indices and spectral types we have gone no further than
dividing the stars in three large groups:O-Type, B-type and A-type.

Those stars that in the $V$ -- $c_{1}$ diagram present a $V$
magnitude much
higher than expected for their spectral type are branded as non-members.
Since the $c_{1}$ index is less affected by extinction than the $(b-y)$ 
color , this is a much
more secure diagnostic than the $V$ -- $(b-y)$ diagram (see 
Figures 4 and 5). The star S3R1N13, which
we have identified as a main-sequence B star from its position in the
$\beta$ -- [$u-b$] diagram, lies away from all other cluster members
in both diagrams and is therefore branded as non-member. Several stars
lying to the right of other cluster members in the $V$ -- $(b-y)$
diagram, grouped together with the members in the
$V$ -- $c_{1}$ diagram. Since NGC~1893 is known to present 
differential reddening, we identify these stars as members with 
particularly high local reddening, something which is confirmed by the 
reddening estimation (see below).

Most stars whose parameters in the $\beta$ -- [$u-b$] diagram
identify as A-type stars present $V$
magnitudes in the $V$ -- $c_{1}$ diagram much
higher than expected for their spectral type and they turn out to be 
non-members. Only four stars lying close to the B9/A0 limit are 
considered likely members (S1R1N4, S2R2N25, S2R3N59 and S4R1N16). 
The list of all the stars that are identified as likely
members is given in Table 8. For these stars we have
calculated their
color excess $E(b-y)$ using the iteration procedure given by Crawford (1978).
These values are also listed in Table 8.

In the $E(b-y)$ -- $V$ diagram (see Figure 6) most cluster 
members concentrate in the $E(b-y) = 0.3 - 0.4$ range, but there are several
stars displaying much higher reddenings. All these stars come from a
relatively small spatial area within the cluster, which must be affected 
by local extinction. Leaving out this group of more highly reddened stars
(including only stars with $E(b-y) < 0.4$),
we determine an average cluster reddening of $E(b-y) = 0.33\pm0.03$ from 
all the other members, where the uncertainty represents the standard
deviation of the mean for the average.
With this value we have calculated the intrinsic colors and magnitudes
of all cluster members, which are listed in Table 8.

For all the stars classified as likely members, we have estimated a
spectral type by using the temperature calibration from
 Napiwotzki et al. (1993). This calibration, based on the [$u-b$]
index, is valid for stars with $T_{\rm eff} \ga 9500\:$K and should
therefore be valid for all members.
By using the expression
\begin{equation}
\Theta \equiv \frac{5040\:{\rm K}}{T_{\rm eff}} = 0.1692 + 0.2828[u-b] -
[u-b]^{2}
\end{equation}
we find the temperatures listed in Table 7. Once we have this information, we derive approximate spectral types by
correlating the estimated $T_{\rm eff}$  with the average
values for each spectral type from Kontizas \& Theodossiou (1980).
All objects are supposed to be unevolved and main-sequence values have been
adopted. 
Of the eight B-type
stars with spectroscopic classification, only three show agreement
with the derived photometric temperature, all the others having
photometric temperatures too low for their spectral type. However,
this effect is not very large and in all cases implies differences of
approximately 1 spectral subtype. Given the magnitude
range covered by our observations, we must suspect that the temperature
calibration is more accurate for later spectral types, because the
photometric temperatures correspond well to the derived absolute
magnitudes, even though there are no spectroscopic spectral types
available for comparison. 
 
\subsection{H--R Diagram}

Figures 7 and 8 show the $M_{V}$ -- $(b-y)_{0}$ 
and $M_{V}$ -- $c_{0}$ diagrams respectively. 
Fabregat \& Torrej\'{o}n (2000) have investigated the different photometric
indices which are commonly used as horizontal axis in the observational HR
diagrams, with regard to the B star region of the main sequence and they conclude
on the basis of four parameters (the span of values taken by the index over 
the B-star range, its intrinsic photometric accuracy, the ratio between
the index variation and the accuracy and how the interstellar reddening affects
the index) that the most efficient way to determine reliable ages for very young
clusters is isochrone fitting to the observational $M_{V}-c_{1}$ HR diagram. 
As can be easily seen, 
the main
sequence is far better defined in the $M_{V}$ -- $c_{0}$ diagram, 
confirming that $c_{0}$ is a much better spectral class indicator than
$(b-y)_{0}$ for early type stars. We have plotted the empirical
ZAMS from Balona
\& Shobbrook (1984) obtained from the relation
$M_{VZAMS}=-2.769+9.599c_{0}-16.207c_{0}^2+17.716c_{0}^3-6.949c_{0}^4$ that
represents analytically the adopted ZAMS in the $M_{V}-c_{0}$ diagram.
Allowing for some uncertainty in the fit to the ZAMS,
we find that the best value of the dereddened distance modulus
is $V_{0} - M_{V} = 13.9\pm0.2$. This value, although higher
than those given by T91 and Fitzsimmons (1993), which are 
13.2 and 13.4 respectively, is compatible within the errors with the
latter, but is not compatible with the value given by
T91. This discrepancy is due mainly to the value of reddening adopted
by Fitzsimmons (1993) being nearer to our value than the value adopted
by T91.  

A second method of determining the distance to the cluster is by using
the procedure from Balona \& Shobbrook (1984) to estimate the absolute 
luminosities of all members from their $\beta$ and $c_{0}$ indexes. 
We find that the values derived
provide rather consistent values for the distance modulus. The average
$V_{0} - M_{V}$ is $13.2\pm0.5$, where the uncertainty is only due
to the dispersion in the estimates and does not include the uncertainties
in the determination of the individual $M_{V}$'s.

By using this procedure, we find two stars (in addition to S3R1N4,
which has been discussed above) which have absolute 
magnitudes which not only give values of $V_{0} - M_{V}$ very far from
the cluster average, but are inconsistent with their spectral types.
These are S4R3N2, which is about one magnitude too bright for the
B2V spectral type given by Hiltner (1966) -- which is perfectly
consistent with the photometric indices -- and S3R3N11, which is
more than one magnitude too faint. Though it is unlikely finding
a field B2V lying just in front of the cluster and at a similar
distance, we brand S4R3N2 as a possible non-member. S3R3N11 has one
of the highest reddenings measured, but it lies very close to cluster
members with similar high reddening. Therefore its membership is also
uncertain. Removing these two stars does not change significantly
the estimate of $V_{0} - M_{V}$.

The values of $V_{0} - M_{V}$ obtained by fitting the ZAMS and by 
averaging the values derived from the individual estimation of 
$M_{V}$ for members are rather different, but compatible within the 
errors (which are much larger for the second method). We prefer
the first determination, not only because of its smaller intrinsic
error, but also because the $M_{V}$'s derived are more consistent
with the spectral types. The four stars with higher effective 
temperature studied by Rolleston et al. (1993) all have
$T_{\rm eff}$ in the range 25000\,--\,27500 K (and these values 
could be an underestimate according to Rolleston et al. 1993). 
Assuming $V_{0} - M_{V} = 13.2$, these stars would all have absolute
magnitudes $ > -3.4$, while taking $V_{0} - M_{V} = 13.9$, they
have  $M_{V}$'s between $-3.3$ and $-4.1$, which correspond well with the
derived spectral types between B0.2V and B0.7V (Vacca et al. 1996).

\subsection{Emission-line candidates}

We find several stars whose $\beta$ indices indicate that they 
could have the H$\beta$ line in emission. Their position in the 
$M_{V}$ -- $c_{0}$ diagram is compatible with their being
Be stars (Marco et al. 2000). These objects, together with
their dereddened indices, have been listed in Table 9.

We performed spectroscopy of all emission-line candidates (except 
for S5001, which is too close to the brighter star S2R1N12 to 
allow separation unless seeing conditions are exceptional) and two
other stars lying nearby in the $M_{V}$ -- $c_{0}$ diagram and having
$\beta$ indices close to the limiting $\beta < 2.55$,
namely S4R2N14 and S2R1N16.

S4R2N15 was found to present no emission lines and a spectrum consistent 
with a K-type main sequence star and it is therefore not a cluster
member. The spectrum of S4R2N14 shows no emission lines either, but
strong H$\alpha$ and H$\beta$ absorption. It is most likely an F 
main-sequence star and therefore not a cluster member.

The three remaining stars (S1R2N23, S2R1N16 and S2R1N26) were all found
to display H$\alpha$ in emission. Their spectra (which are presented
in Fig. 9 have low signal-to-noise
ratios due to their faintness and do not cover the classification region.
However, comparison with the stars in the digital atlas of Jacoby et al.
(1984) indicates that they are all approximately F-type stars.

The earliest spectrum is that of S1R2N23, which is close to F0. This star
also displays the strongest H$\alpha$ among the group, with an Equivalent
Width (EW) of $-18$\AA. This line, as well as H$\beta$, probably displays
inverse P-Cygni profiles, indicative of mass inflow, which are typically 
seen in many young stars. Also in emission is the Ca\,{\sc ii}  $\lambda\lambda$ 8498, 8542, 8662 \AA\ triplet. Measurements in previous studies
(Hoag 1961; Cuffey 1973; Moffat \& Vogt 1975; Massey et al. 1995) indicate
strong variability, specially in the $(U-B)$ color.

The spectrum of S2R1N16 is that of a mid-F star and its H$\alpha$ emission
is rather weaker (EW=$-7$\AA). The infrared Ca\,{\sc ii} triplet is in 
absorption and there is no evidence for other strong emission lines.
Measurements in previous studies indicate very strong variability.

The latest spectrum is that of S2R1N26, which is  very late F or early G.
Its H$\alpha$ emission is also the weakest, with an EW=$-5$\AA. There is
no evidence for any other emission lines. Previous photometric measurements 
are consistent with little variability.

The presence of emission lines identifies all three objects as young
(pre-Main Sequence) stars. They are probably massive T Tauri
stars, with S1R2N23 coming within the group of Herbig AeBe stars according
to the classification of Th\'{e} et al. (1994), and they are all likely 
members of NGC~1893. Their intrinsic luminosities
must be therefore much higher than those of main-sequence stars of the same
spectral type.

\subsection{Previously cataloged Be stars}

We also observed the two NGC 1893 stars cataloged by Massey et al. (1995)
as Be stars, namely S1R2N35 and S3R1N3. Their spectra are presented
in Fig. 10. Both stars display very strong emission in
H$\alpha$ and H$\beta$, as well as in the upper Paschen lines (Pa14 to
Pa17 are clearly seen in both stars). The O\,{\sc i} $\lambda$7773 \AA\ and 
$\lambda$8446 \AA\ lines are also prominent. The yellow and red continuum
is dominated by many strong permitted Fe\, {\sc ii} emission lines, among which
the complex dominated by Fe\, {\sc ii} $\lambda\lambda$ 5169, 5198, 5235,
5276 \AA\ is particularly strong in S3R1N3.

The strength of the H$\alpha$ emission line (EW$=-71$\AA) and the overall
emission spectrum is compatible with an extreme classical Be star, but
more typical of a Herbig Be young stellar object. The photometric 
indices of this star
indicate that this object has a spectral type close to B0. Therefore,
with $V=13.6$, it is much fainter than expected (for example, it is 2.3
mag fainter than the B0.2V star S3R1N17). This could be indicating
that it is surrounded by a local nebulosity and it is indeed a 
pre-main-sequence Herbig Be star. Classical Be stars tend on average to 
be brighter than corresponds to their spectral type (Fabregat \& 
Torrej\'{o}n 1998).

The emission spectrum of S1R2N35 is still stronger, with the EW of H$\alpha$
amounting to $-88$ \AA. Our blue spectrum of the source shows red-shifted 
emission components in H$\gamma$ and H$\delta$. The spectral type is
$\approx$ B0 and again the star is at least one magnitude too faint for the
spectral type, indicating the presence of substantial local extinction.
Therefore this star is most likely a Herbig Be object as well. References
in the literature indicate that this star has brightened from $V\sim12.8$
(Hoag 1961; Moffat \& Vogt 1975) to  $V\sim12.3$ ( Fitzsimmons 
1993; Massey et al. 1995).

Schild \& Romanishin (1976) list a third Be star in NGC 1983, which has
been subsequently identified with S2R2N43 in the Webda database
(Mermilliod 1999). However, we do not believe that this identification is
correct. Massey et al. (1995) found no evidence of
emission from S2R2N43. Rolleston et al. (1993) selected this star
as a typical main-sequence early-B star from the photometric data
of Fitzsimmons (1993) and studied its chemical composition using
high-resolution spectroscopy without finding any indication of emission.

Our spectra of this star offer no evidence of emission
in the whole optical range. We
find a spectral type of B0.5V for S2R2N43, rather than the B1III 
derived by Massey et al. (1995). This spectral type is in agreement
with the stellar parameters derived by  Rolleston et al. (1993). 
Moreover, assuming that the reddening to the star
is the average for the cluster, our measured $V=11.5$ implies an absolute
magnitude $M_{V}=-4.0$, which is that expected for the spectral type
(Vacca et al. 1996). Therefore if there is another emission line B star 
in the cluster, it is not S2R2N43.

\section{Discussion}

We find a distance modulus $V_{0} - M_{V} = 13.9\pm0.2$, moderately
larger than that of previous authors. T91 give 
a rather lower $V_{0} - M_{V} = 13.2\pm0.1$. There are two main
contributions to this discrepancy. On the one hand, our derived 
$E(b-y) = 0.33\pm0.03$ implies an extinction $A_{V} = 1.42\pm0.13$,
compared to the $A_{V} = 1.68\pm0.06$ from T91. Our
determination of the reddening is based on a much larger sample of 
stars (but still leaving out all high-reddening members and all
O-type stars). 
If the reddening of all member stars (including those with higher
reddening, but leaving out O-type stars) is averaged, then a value
$E(b-y) = 0.37\pm0.09$, much closer to that from T91
is obtained. However, because of the spatial 
concentration of the highly reddened members and the obvious gap
between them and the rest of the cluster in the 
$E(b-y)$ -- $V$ diagram (see Figure 6), we are 
confident that the
lower value $E(b-y) = 0.33\pm0.03$ is a more accurate measurement
of the true interstellar reddening to the cluster.
 
In addition, most of the discrepancy in the  distance modulus estimate
comes from the
fact that a higher value is needed in order to fit the ZAMS to the 
observational data. Our data include a much larger number of members than
the work by T91 and Fitzsimmons (1993). Stars in the
mid- and late-B range were very scarce in previous work, allowing for
a looser fit of the ZAMS. As can be seen in Figure 8, the
ZAMS from Balona \& Shobbrook provides a very good lower envelope to 
stars from B1 to late B, but lies above the earliest B-type stars. 
It could be argued that the ZAMS should be fit to these B0-type stars,
which would mean increasing the distance modules by an extra 
$\sim 0.5\:$mag. However, this would mean that {\em all} the stars
later than $\sim$ B1 have not yet reached the ZAMS, implying an age
for the cluster of less than 1$\:$Myr, which is in contradiction with
the fact that at least some of the O-type stars seem to be close to
the end of the main-sequence. The weak He\, {\sc ii}~$\lambda4686$\AA\ line
in S3R2N15 has an EW very close to that considered the limit between
main-sequence and giant stars (Mathys 1988), which would set the age
of this star at $\approx 4$ Myr for an assumed $M_{*}\geq 35 M_{\sun}$
at O5V (Vacca et al. 1996).

In any case, the presence of very massive Herbig Be stars indicates that
star formation is still active in this cluster. Since stars in the mid-B
range at least seem to have reached the main sequence, this should be
indicating that star formation has been going on for some time in NGC~1893.
Further observations are needed in order to explore the possible 
existence of a larger pre-main-sequence population. The observed 
spectral distribution of pre-MS stars in NGC~1893 is indeed unexpected, 
suggesting that not all the pre-MS objects in the cluster have been
observed. Moreover, since at least one O-type star seems to have an
age $\approx 4$ Myr, which is approximately the age assumed for the
whole cluster by previous authors, good determinations of the ages of
all the O stars are required in order to explore at which stage the
most massive stars formed. Since massive stars are still forming, it
may seem that massive stars can form both at the beginning or
at end of the epoch of star formation. Knowing whether they can form
at any given time will need in-depth study of the ages of all O-type
stars in NGC~1893.

The identification of all emission-line objects as pre-MS stars means that 
there are no classical Be stars in NGC~1893. This is in agreement with the
suggestion by Fabregat \& Torrej\'{o}n (2000) that there are no classical
Be stars in clusters in which star formation is still active.

\section{Conclusions}

We have identified $\approx 50$ very likely members of NGC~1893 and
therefore we have a sample of likely members larger than
previous studies ($\sim$ 24 members in
T91 and $\sim$ 30 members in Fitzsimmons 1993) to determine the
cluster astrophysical parameters.

We derive a reddening to the cluster
$E(b-y)=0.33\pm0.03$. This value is
similar to that obtained by Fitzsimmons (1993) but is incompatible
with the value adopted by T91. We show that the reddening
is variable across the face of the cluster and can reach much higher
values than the average in some areas. If we include the highest
values of individual reddenings in the calculation of the average,
then our value would be rather closer to the value given by T91. 
We have not adopted this value, because all the high values belong to
stars coming from a relatively small spatial area within the cluster,
which must be affected by local extinction.

We derive a distance modulus of $13.9\pm0.2$ by fitting the ZAMS from
Balona \& Shobbrook (1984).
We obtain a higher value than the  13.2 obtained by T91 and
  the 13.4 by Fitzsimmons (1993), both based on a smaller number of
  likely members.  

Five emission-line stars are shown to be pre-main-sequence objects,
extending from about B0 to very late F spectral type. This suggests 
that several other pre-main-sequence objects could be present in this cluster,
specially at fainter magnitudes than those observed by us. Further 
photometric and spectroscopic work must be conducted to confirm
this suggestion.

\acknowledgments

The JKT and INT are operated on the island of La Palma by the Royal Greenwich 
Observatory in the Spanish Observatorio del Roque de Los Muchachos of 
the Instituto de Astrof\'{\i}sica de Canarias. The G.D. Cassini telescope
is operated at the Loiano Observatory by the Osservatorio Astronomico di
Bologna. We are thankful to all the staff at the Loiano Observatory for
their support with the observations. This research has made use of data
from the ING Archive. 
We would like to thank the Spanish CAT panel for allocating observing time
to this project. AM would like to thank Dr.~J.~Fabregat for his help with the
observations. IN acknowledges receipt of an ESA external research
fellowship.

\clearpage

\begin{figure}
\begin{center}
\begin{tabular}{cc}
\psfig{file=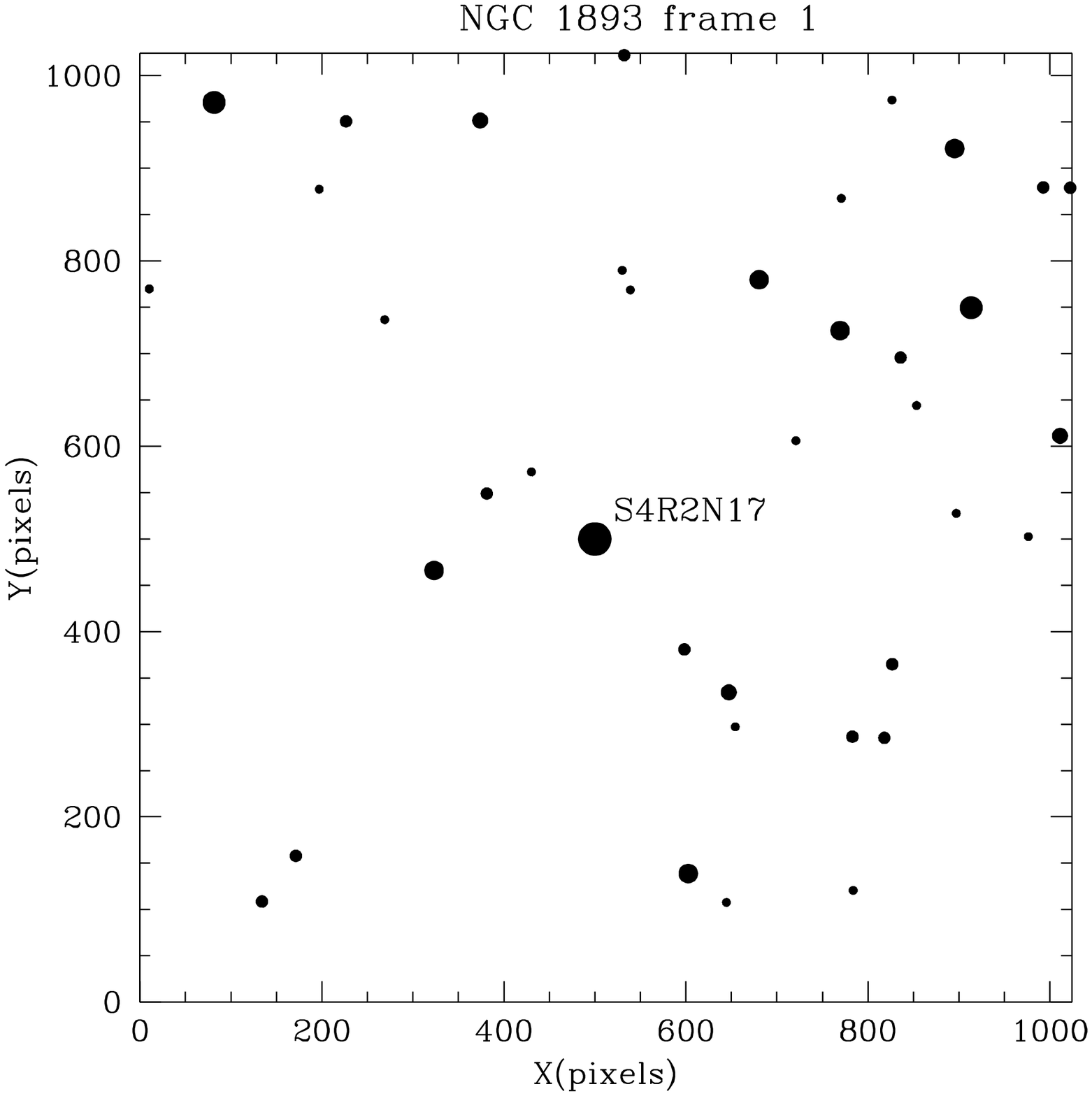,width=8cm,height=8cm} & \psfig{file=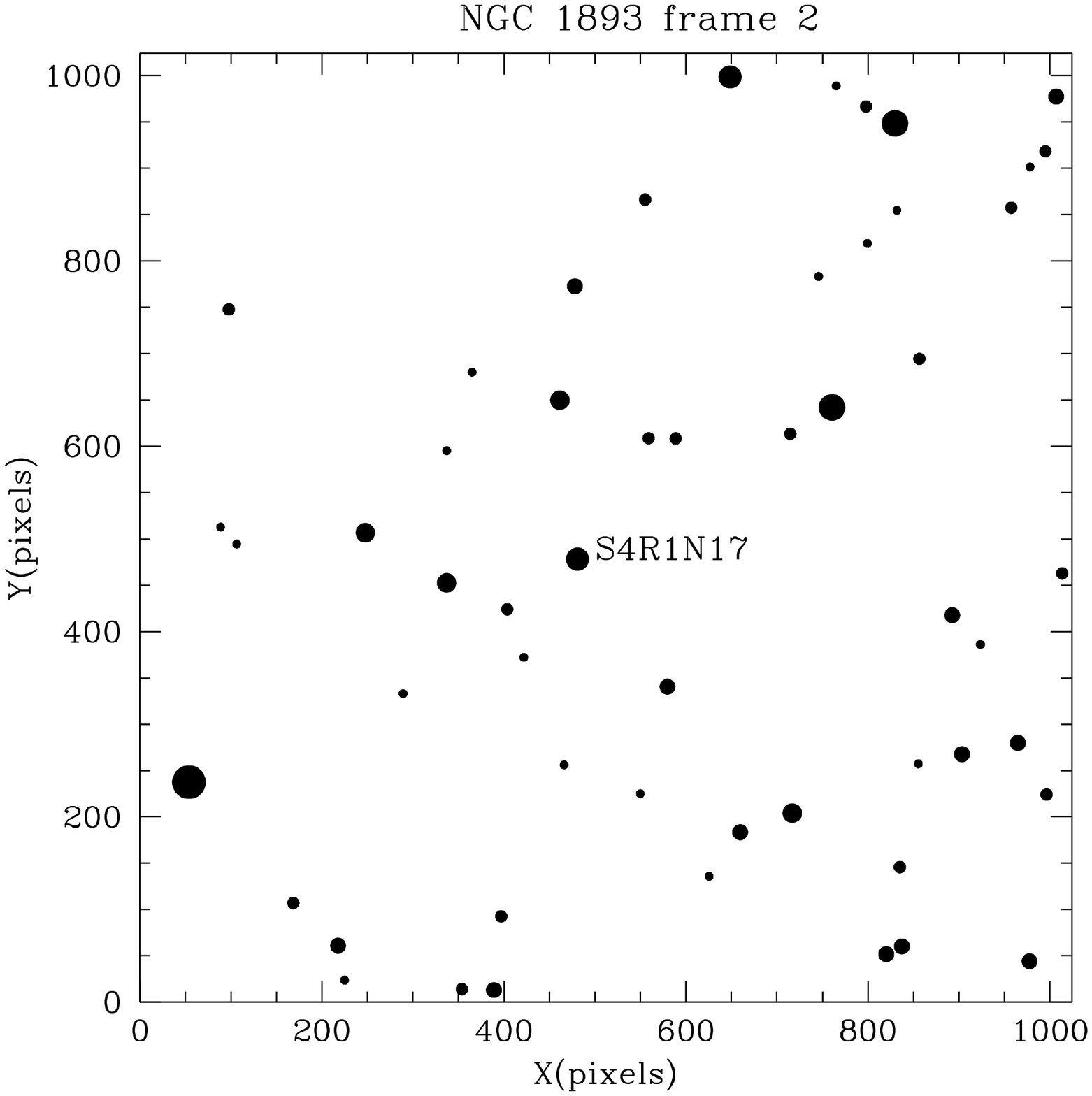,width=8cm,height=8cm} \\
\psfig{file=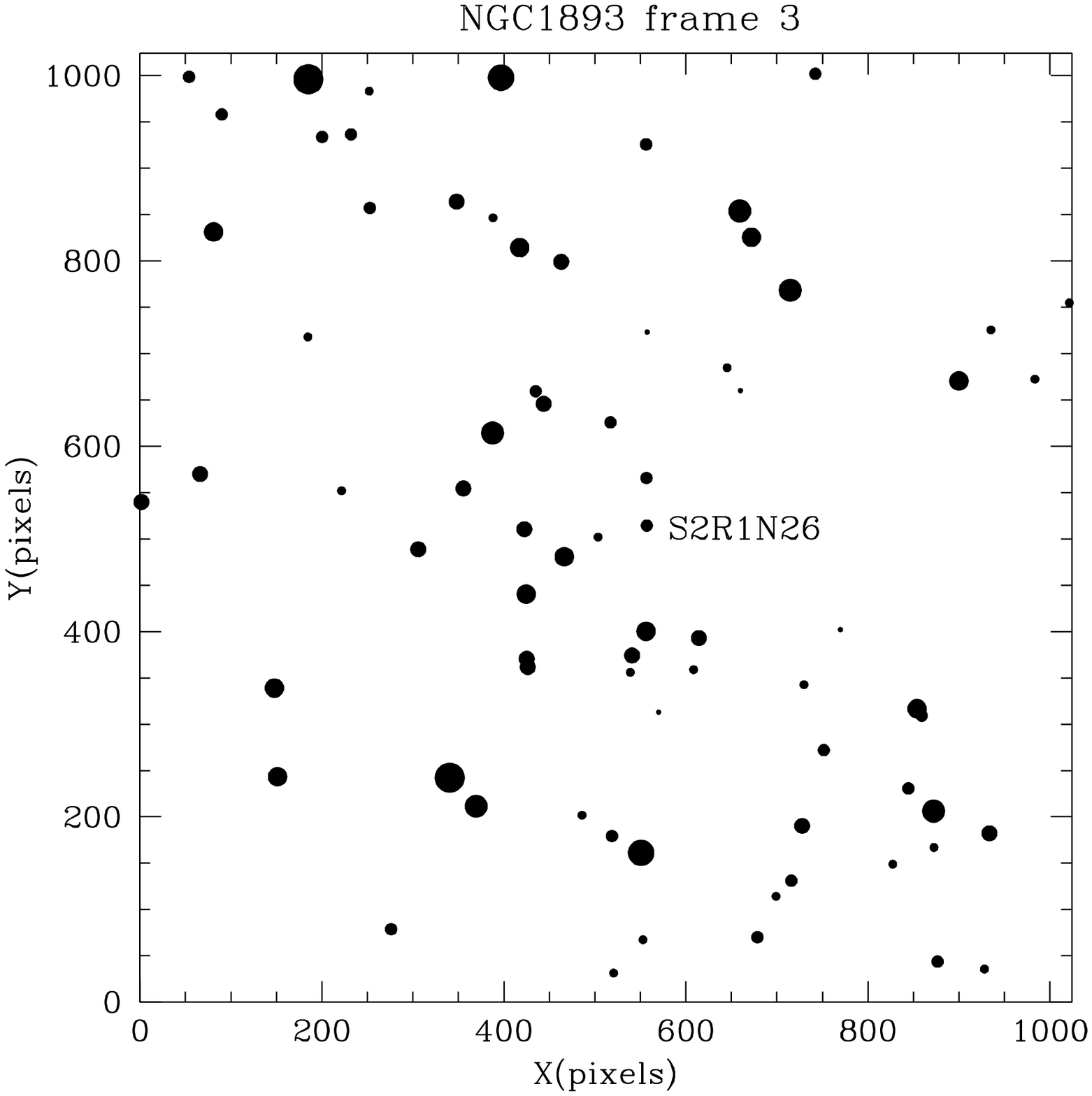,width=8cm,height=8cm} & \psfig{file=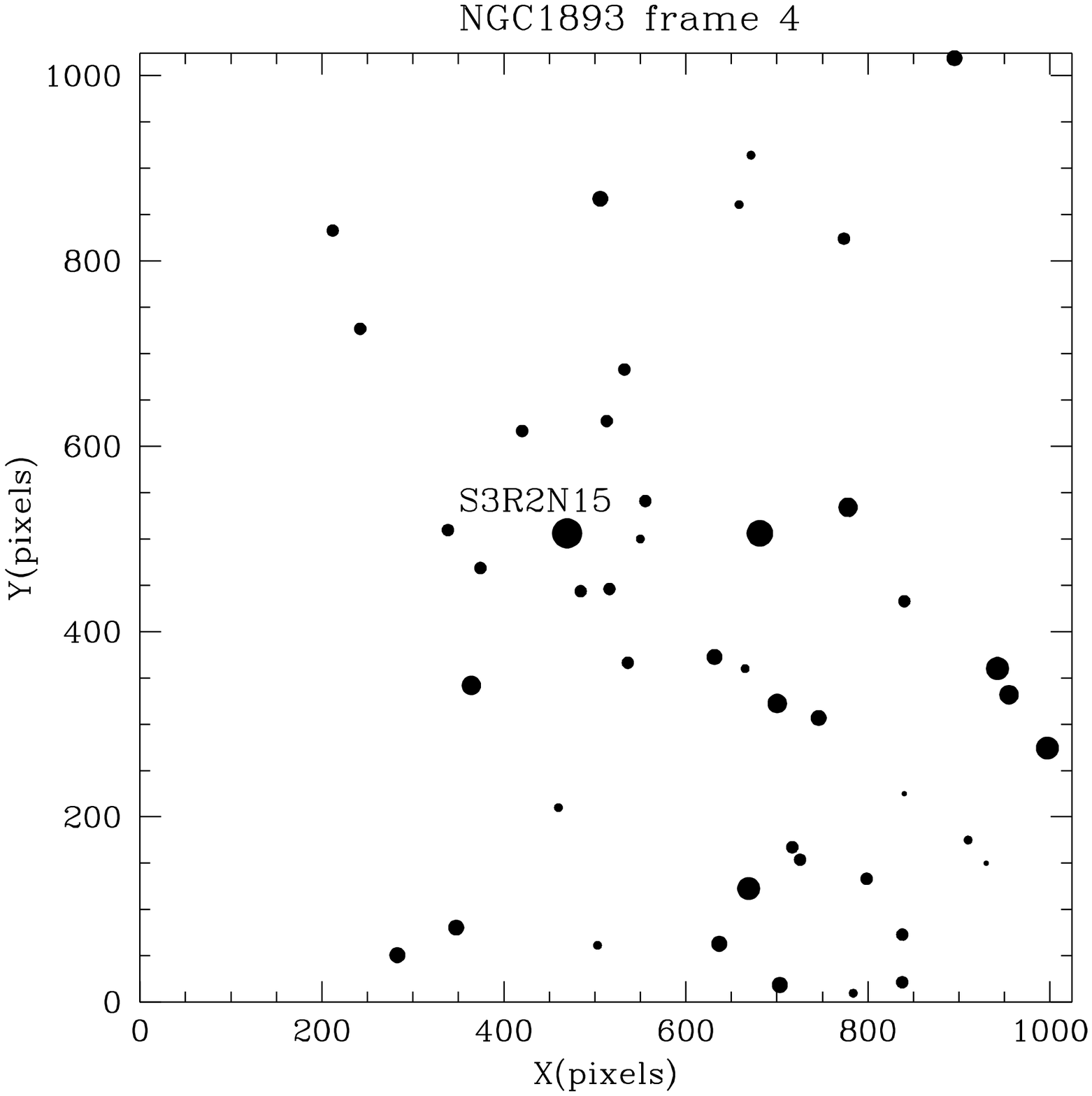,width=8cm,height=8cm} 
\end{tabular}
\end{center}
\caption{(a)-(d). Schematic maps of the four region observed in NGC 1893.
The size of the dots represents relative brightness of stars in the field.
North is down and East left in all fields. \label{fig1}}
\end{figure}

\clearpage
\begin{figure}
\plotone{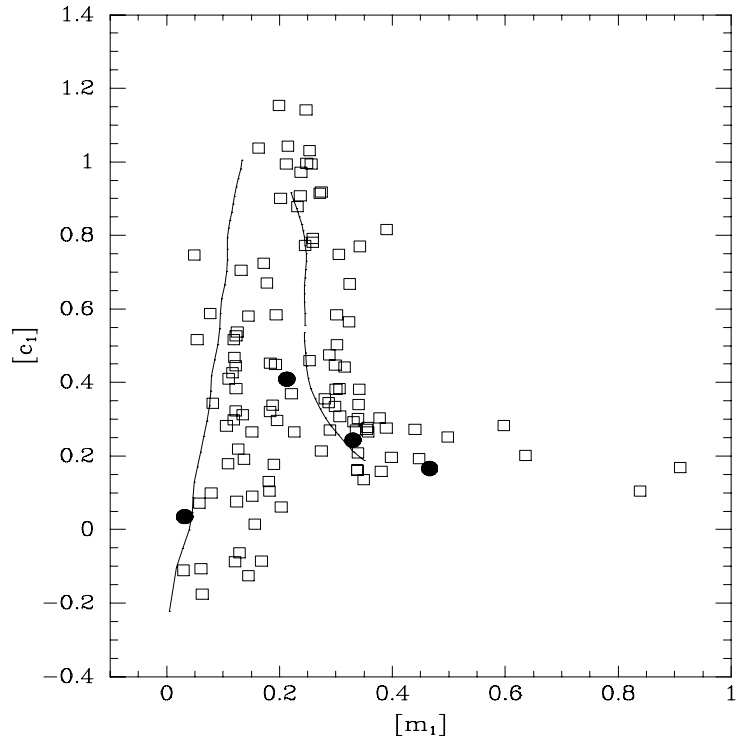}
\caption{The [$c_{1}$] -- [$m_{1}$] diagram for all the stars
observed in the field of NGC~1893. The thin solid lines represent
the average loci of field B-type (left) and A-type (right) main sequence
stars. Filled circles represent stars with H$\beta$ in
emission. \label{fig2}}
\end{figure}

\clearpage
\begin{figure}
\plotone{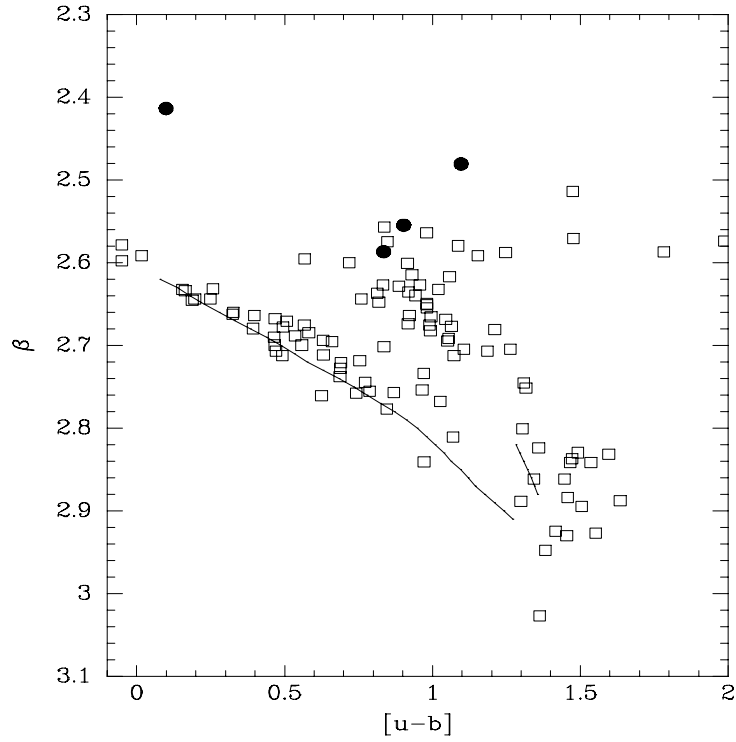}
\caption{The $\beta$ -- [$u-b$] diagram for all the stars
observed in the field of NGC~1893. The thin lines represent the
loci of main-sequence B stars (long line) and main-sequence A1--A2 
stars (short line). The turnover in the $\beta$ -- [$u-b$] correlation
occurs approximately at spectral type A0. Filled circles represent
stars with H$\beta$ in emission. \label{fig3}}
\end{figure}

\clearpage
\begin{figure}
\plotone{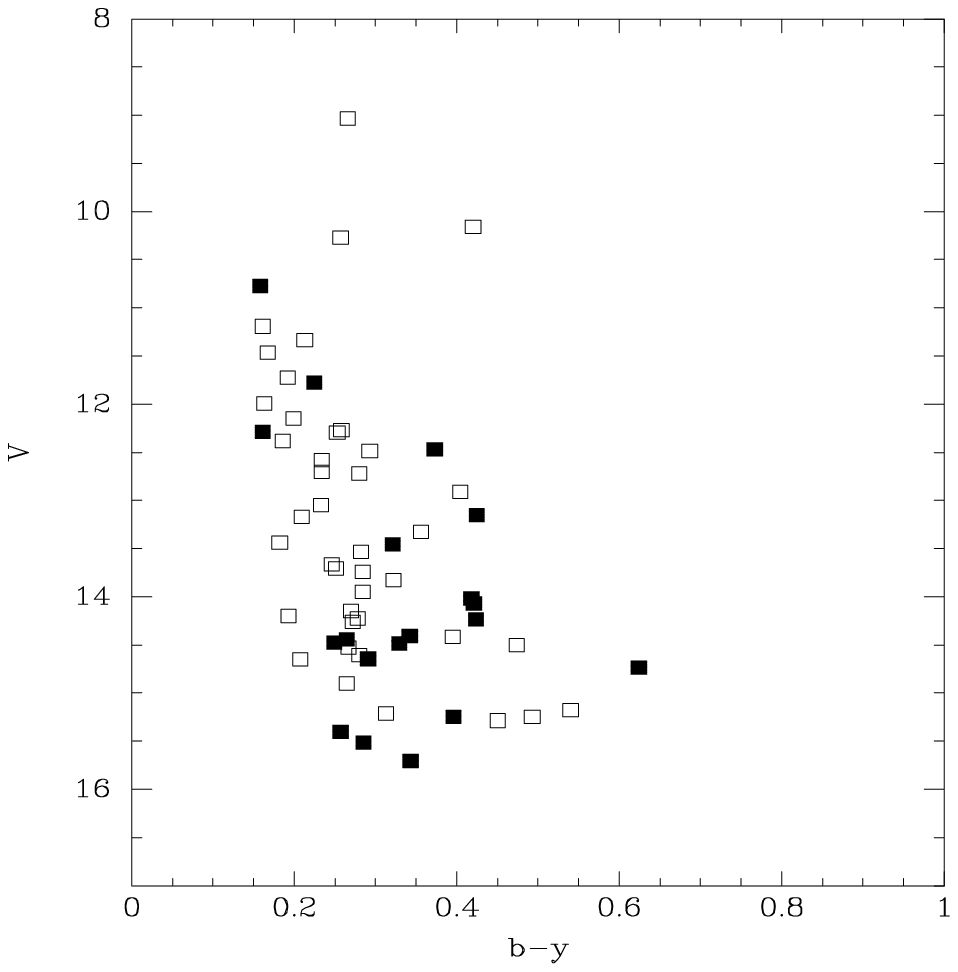}
\caption{$V$ magnitude against $(b-y)$ color for suspected members of NGC~1893. Filled squares represent stars that have finally been classed 
as non-members, while open squares represent those that have been considered
very likely members. \label{fig4}}
\end{figure}

\clearpage
\begin{figure}
\plotone{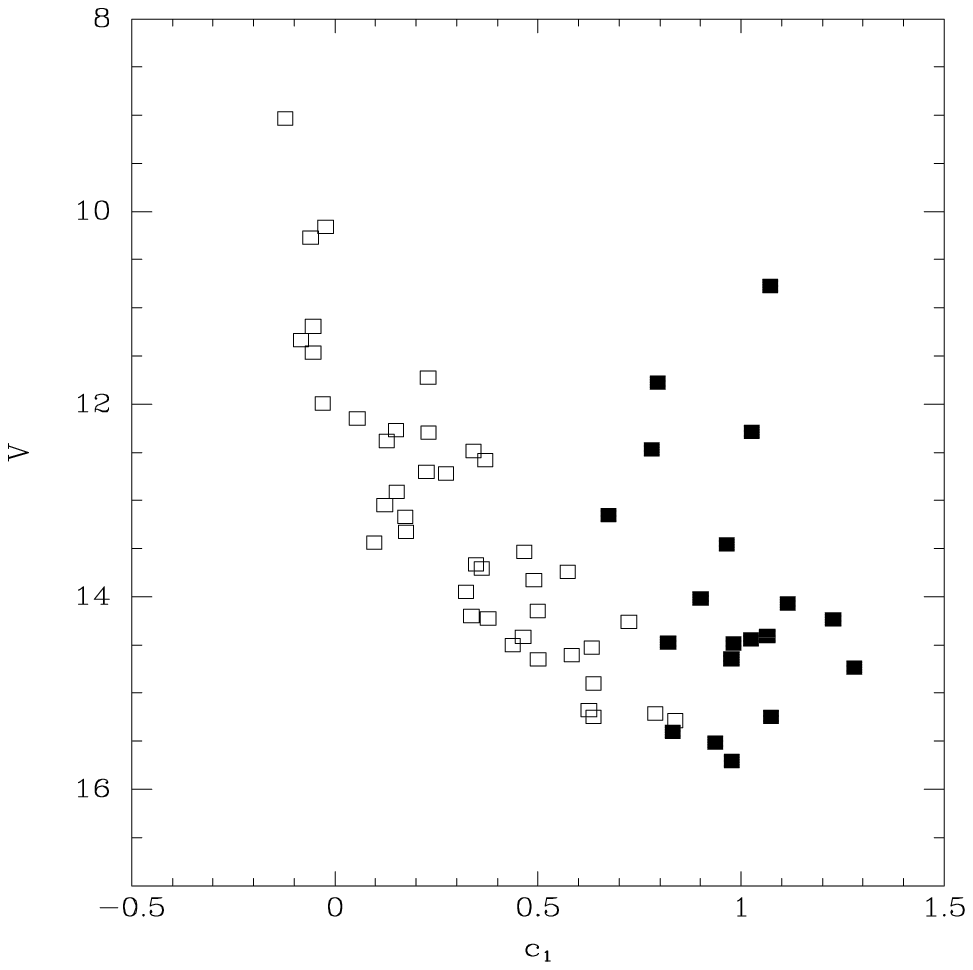}
\caption{$V$ magnitude against $c_{1}$ index for suspected members of NGC~1893. Filled squares represent stars that have finally been classed 
as non-members, while open squares represent those that have been considered
very likely members. \label{fig5}}
\end{figure}

\clearpage
\begin{figure}
\plotone{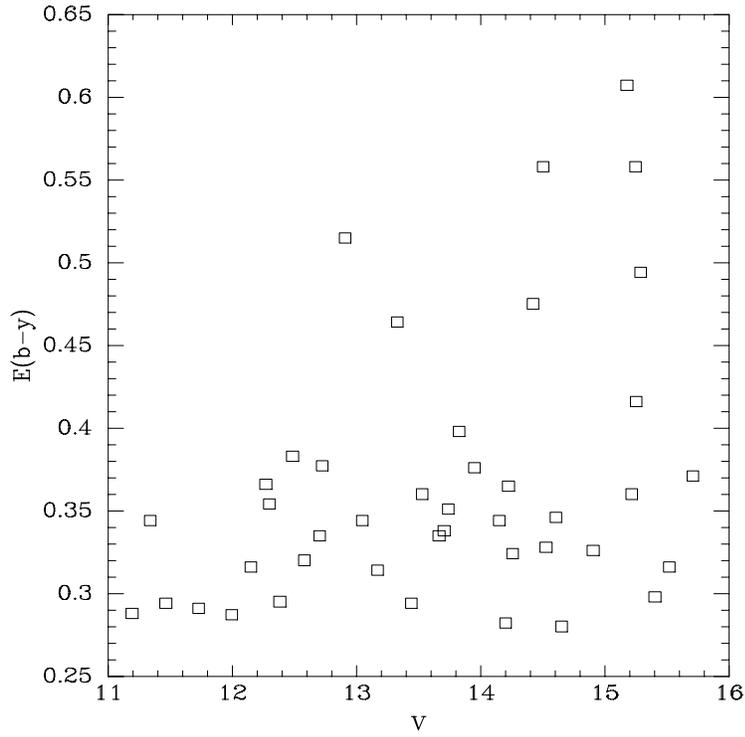}
\caption{Individual color excess $E(b-y)$, calculated by applying Crawford's (1978) procedure, against apparent visual magnitude
 for all likely members of NGC~1893. \label{fig6}}
\end{figure}

\clearpage
\begin{figure}
\plotone{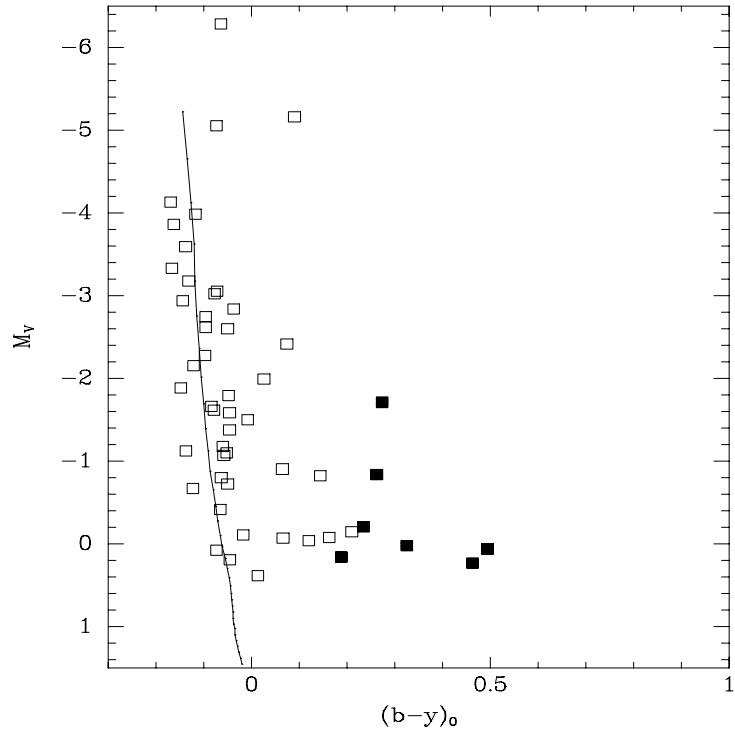}
\caption{Absolute magnitude $M_{V}$ (calculated using the average distance modulus for the cluster) against $(b-y)_{0}$ color. Filled squares represent stars studied spectroscopically. The solid line represents the theoretical ZAMS from Perry et al. (1987). \label{fig7}}
\end{figure}

\clearpage
\begin{figure}
\plotone{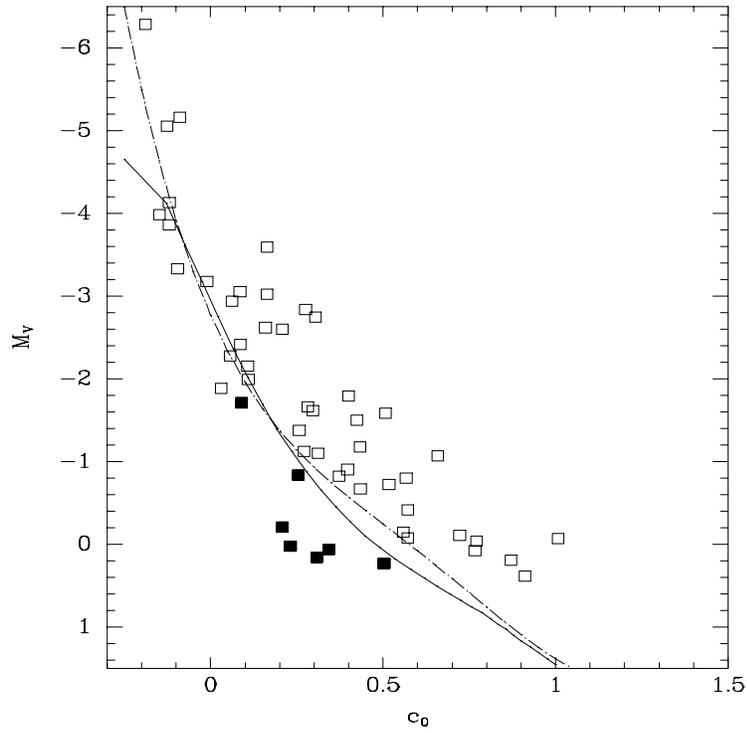}
\caption{Absolute magnitude $M_{V}$ (calculated using the average distance modulus for the cluster) against $c_{0}$ index. Filled squares represent stars studied spectroscopically. The solid line represents the theoretical
ZAMS from Perry et al. (1987). The dashed line represents the theoretical
ZAMS from Balona \& Shobbrook (1984). \label{fig8}}
\end{figure}

\clearpage
\begin{figure}
\plotone{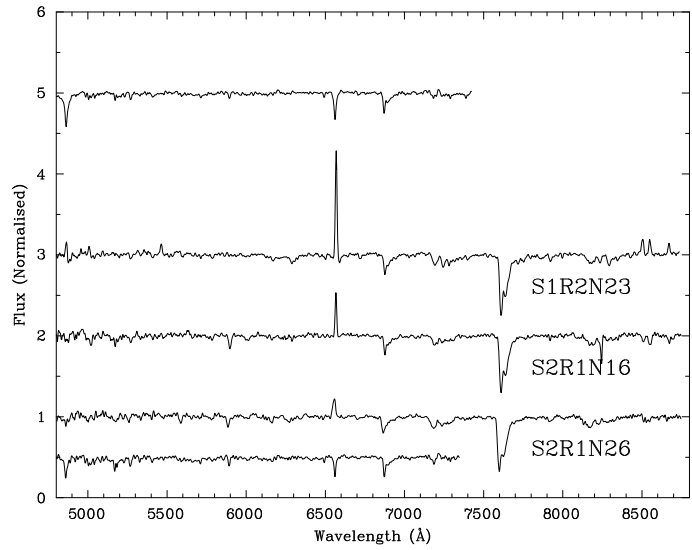}
\caption{Low resolution spectra of the three candidate emission-line stars which were actually found to display H$\alpha$ in emission. Also
shown for comparison are the spectra of HD~10032 (F0V, top) and HD~31084 
(F9V, bottom) from the electronic atlas of Jacoby et al. (1984), binned to the same resolution. All the spectra have been divided by a spline fit to the continuum for normalisation and arbitrarily offset for display. \label{fig9}}
\end{figure}

\clearpage
\begin{figure}
\plotone{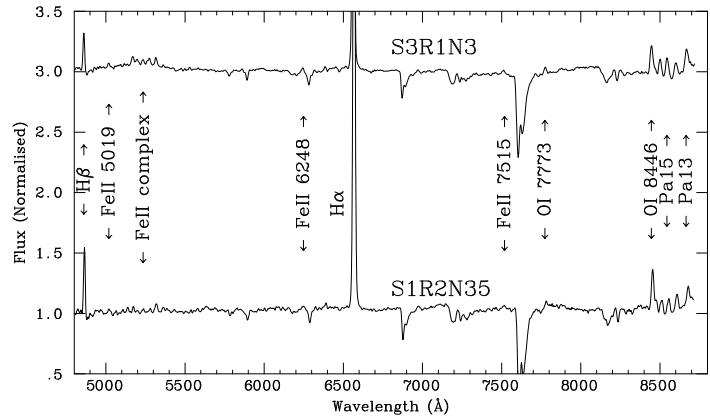}
\caption{Low resolution spectra of the two stars in NGC 1893 previously cataloged as Be stars, S1R2N35 and S3R1N3. The spectra have 
been divided by a spline fit to the continuum for normalisation. \label{fig10}  }
\end{figure}

\clearpage
\begin{figure}
\plotone{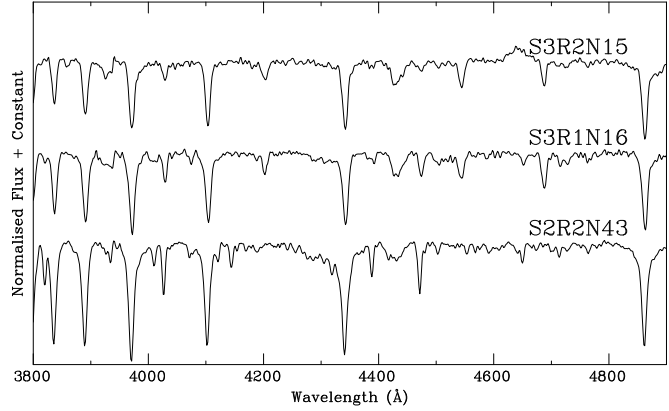}
\caption{Spectra of bright stars in NGC 1893 for which our spectral classification differs from previous estimates. In S3R2N1 (top), the very
weak He\, {\sc i}~$\lambda 4471$\AA, together with
He\, {\sc i}+ He\, {\sc ii}~$\lambda 4026$\AA$\,\simeq\,$He\, {\sc ii}~$\lambda
4200$\AA, makes the star earlier than O6V. The quantitative criterion
of Mathys (1988) places the star on the limit between O5V and O5.5V. The
weakness of He\, {\sc ii}~$\lambda 4686$\AA\ and strength of N\, {\sc iii}
emission point to the star's being slightly evolved from the main
sequence. In S3R1N16 (middle), the condition
He\, {\sc i}~$\lambda 4471$\AA$\,\simeq\,$He\, {\sc ii}~$\lambda 4541$\AA\
defines O7V. In S2R2N43, the weak Si lines mark the star as main
sequence, while the strength of the O\, {\sc ii} spectrum indicates B0.5V.}
\end{figure}

\clearpage

\begin{deluxetable}{lcc}
\tablewidth{0pt}
\tablecaption{Observational details of the photometry. Observations 
were taken using the  1.0-m JKT on December 11th and December 22nd
1997 for four frames.\label{tb1}}
\tablehead{
\colhead{}&\colhead{Central Star}&\colhead{Coordinates (1950)}}
\startdata
\#1&S4R2N17&$\alpha$=5h 19m 12.00s $\delta$=$+33\degr$ $28\farcm$ $00\farcs4$\\
\#2&S4R1N17&$\alpha$=5h 19m 23.24s $\delta$=$+33\degr$ $26\farcm$ $40\farcs0$\\
\#3&S2R1N26&$\alpha$=5h 19m 31.09s $\delta$=$+33\degr$ $22\farcm$ $11\farcs3$\\
\#4&S3R2N15&$\alpha$=5h 19m 22.63s $\delta$=$+33\degr$ $19\farcm$ $29\farcs0$\\
\enddata
\end{deluxetable}

\clearpage

\begin{deluxetable}{crrrrrc}
\tablewidth{0pt}
\tablecaption{Standard stars with their cataloged values and spectral types taken from
literature.}
\tablehead{
\colhead{Number}&\colhead{$V$}&\colhead{$(b-y)$}&\colhead{$m_{1}$}&\colhead{$c_{1}$}&\colhead{$\beta$}&\colhead{Spectral Type}
}
\startdata
\cutinhead{h Persei}
837&14.080&0.393&0.000&0.918&2.801&          \\
843& 9.320&0.277&-0.050&0.166&     &B1.5V     \\
867&10.510&0.393&0.161&0.375&2.613&          \\
869&      &     &     &     &2.700&          \\
935&14.020&0.362&-0.004&0.854&2.802&          \\
950&11.290&0.318&-0.048&0.214&2.642&B2V       \\
960&      &     &     &     &2.767&          \\
978&10.590&0.305&-0.039&0.177&2.643&B2V - B1.5V\\
982&      &     &     &     &2.796&          \\
1015&10.570&0.225&0.033&0.741&     &B8V       \\
1078& 9.750&0.316&-0.065&0.167&2.610&B1V - B1Vn\\
1181&12.650&0.372&-0.034&0.379&2.718&          \\
\cutinhead{$\chi$ Persei}
2133&      &     &     &     &2.676&          \\
2139&11.380&0.255&-0.033&0.196&2.649&B2V       \\
2147&14.340&0.406&-0.050&1.002&2.863&          \\
2167&13.360&0.352&-0.056&0.627&2.752&          \\
2185&10.920&0.283&-0.049&0.406&2.700&B2Vn      \\
2196&11.570&0.250&-0.006&0.210&2.670&B1.5V     \\
2200&      &     &      &     &2.721&          \\
2232&11.110&0.292&-0.105&0.207&2.651&B2V       \\
2235& 9.360&0.316&-0.088&0.150&2.611&B1V       \\
2251&11.560&0.302&-0.042&0.349&2.709&B3V       \\
\cutinhead{NGC 2169}
11&10.600&0.084&0.065&0.541&2.698&B8V       \\
15&11.080&0.130&0.109&0.944&2.864&B9.5V     \\
18&11.800&0.115&0.105&0.912&2.872&B9.5V     \\
\cutinhead{NGC 6910}
7&10.360&0.670&-0.160&0.110&2.612&B0.5V     \\
11&10.900&0.770&0.420&0.430&2.555&          \\
13&11.720&0.660&-0.140&0.220&2.647&B1V       \\
14&11.730&0.590&-0.100&0.220&2.652&B1V       \\
15&12.220&0.590&-0.110&0.330&2.679&          \\
17&12.660&0.670&-0.120&0.310&2.659&          \\
18&12.810&0.750&-0.140&0.290&2.680&          \\
19&12.920&0.640&-0.120&0.380&2.662&          \\
20&12.980&0.610&-0.130&0.420&2.692&          \\
\cutinhead{NGC 1039}
92&11.960&0.303&0.138&0.481&2.678&          \\
96& 9.740&0.086&0.176&0.973&2.890&          \\
97&11.820&0.144&0.198&0.900&2.855&          \\
102&10.760&0.151&0.194&0.894&2.848&          \\
105&11.220&0.176&0.204&0.796&2.817&          \\
109&10.030&0.066&0.152&1.013&2.916&          \\
111& 9.950&0.055&0.163&1.021&2.908&          \\
\enddata
\end{deluxetable}

\clearpage

\begin{deluxetable}{lrrrrc@{}crrrrrrrrr}
\tabletypesize{\scriptsize}
\tablewidth{0pt}
\tablecaption{Catalog of 41 standard stars observed and transformed to the Crawford-Barnes $uvby$ and the Crawford-Mander $H\beta$ standard systems. The internal rms errors of the mean measure in the transformation of each star are given in columns 8 to 11 in units of 0.001 mag. Columns 12 to 16 give the difference $\Delta$=standard value minus transformed value, in units of 0.001 mag. N is the number of measures of each standard star in the transformation in $V$, $(b-y)$, $m_{1}$, $c_{1}$. The number of measures in $\beta$ is one for each standard star in the transformation.}
\tablehead{
\colhead{Star}&\colhead{$V$}&\colhead{$(b-y)$}&
\colhead{$m_{1}$}&\colhead{$c_{1}$}&\colhead{$\beta$}&\colhead{$N_{uvby}$}&
\colhead{$\sigma_{V}$}&\colhead{$\sigma_{(b-y)}$}&
\colhead{$\sigma_{m_{1}}$}&\colhead{$\sigma_{c_{1}}$}&
\colhead{$\Delta_{V}$}&\colhead{$\Delta_{(b-y)}$}&\colhead{$\Delta_{m_{1}}$}&
\colhead{$\Delta_{c_{1}}$}&\colhead{$\Delta_{\beta}$}
}
\startdata
\cutinhead{h Persei}
869&-&-&-&-&2.724&-&-&-&-&-&-&-&-&-&-024\\
837& 14.095&0.411&-0.031&0.907&2.789& 3&006&009&024& 024&-016&-019& 032&011&012\\
843&  9.330&0.319&-0.151&0.240&-& 1&  -   &  -   &  -   &  -    &-010&-042& 101&-074&-\\
867& 10.572&0.376& 0.177&0.379&2.666& 5&006&008&013& 025&-063& 016&-015&-005&-053\\
935& 14.051&0.411&-0.076&0.856&2.781& 6&018&007&018& 038&-031&-049& 073&-003&021\\
950& 11.297&0.337&-0.079&0.221&2.630& 1&  -   &  -   &  -   &  -  &-007&-019&031&-007&012\\
960&-&-&-&-&2.727&-&-&-&-&-&-&-&-&-&040\\
978& 10.646&0.328&-0.070&0.194&2.634& 5&016&013&016&023&-056&-023& 031&-017&009\\
982&-&-&-&-&2.779&-&-&-&-&-&-&-&-&-&017\\
1015& 10.573&0.223& 0.029&0.677&-& 2&037&023&006& 030&-003& 003& 005&064&-\\
1078&  9.775&0.317&-0.043&0.138&2.611& 3&008&021&042& 036&-025&-002&-021&030&-001\\
1181& 12.655&0.352&-0.008&0.338&2.703& 6&022&022&044& 030&-006& 020&-025&041&015\\
\cutinhead{$\chi$ Persei}
2133&-&-&-&-&2.658&-&-&-&-&-&-&-&-&-&018\\
2139& 11.351&0.298&-0.097&0.235&2.641& 2&011&024&030& 007&029&-043& 064&-039&008\\
2147& 14.359&0.392&-0.083&1.022&2.783& 5&023&034&051& 051&-019& 013& 033&-020&080\\
2167& 13.364&0.347&-0.080&0.616&2.733& 6&012&008&012& 024&-004& 005& 025&011&019\\
2185& 10.926&0.275&-0.018&0.412&2.688& 3&008&013&038& 044&-006& 008&-031&-006&012\\
2196& 11.549&0.304&-0.066&0.246&2.632& 6&012&013&023&023&021&-052& 056&-034&038\\
2200&-&-&-&-&2.707&-&-&-&-&-&-&-&-&-&014\\
2232& 11.052&0.238&-0.029&0.177&2.639& 4&013&013&041& 035&058& 054&-077&030&012\\
2235&  9.365&0.311&-0.071&0.131&2.575& 3&012&015&046& 039&-005& 005&-016&018&036\\
2251& 11.563&0.297&-0.028&0.367&2.689& 6&009&011&027& 030&-004& 004&-013&-018&020\\
\cutinhead{NGC 2169}
11& 10.538&0.076& 0.061&0.545&2.719& 1&-     &-     &   -  &  -   &062& 008& 004&-004&-021\\
15& 11.023&0.122& 0.092&0.939&2.856& 1&-     &   -  &   -  & -    &057& 008& 017&005&008\\
18& 11.733&0.053& 0.260&0.813&2.883& 1&-     &-     & -    &   -  &067& 062&-155&099&-011\\
\cutinhead{NGC 6910}
7& 10.320&0.667&-0.150&0.092&2.652& 2&011&004&012& 016&041& 003&-011&018&-040\\
11&-&-&-&-&2.639&-&-&-&-&-&-&-&-&-&-084\\
13& 11.681&0.619&-0.082&0.211&2.661& 2&004&023&038& 004&039& 041&-058&009&-014\\
14& 11.730&0.569&-0.053&0.201&2.678& 2&004&028&040& 018&001& 021&-048&019&-026\\
15& 12.193&0.609&-0.134&0.359&2.698& 2&007&006&014& 002&027&-019& 024&-029&-019\\
17& 12.635&0.700&-0.149&0.294&2.688& 2&008&010&025& 040&025&-030& 029&017&-029\\
18& 12.816&0.755&-0.132&0.296&2.705& 2&014&004&006& 048&-006&-005&-009&-006&-025\\
19& 12.897&0.610&-0.065&0.402&2.709& 2&025&035&037& 047&023& 031&-056&-022&-047\\
20& 12.920&0.619&-0.118&0.446&2.730& 2&016&001&019& 006&060&-009&-013&-026&-038\\
\cutinhead{NGC 1039}
92& 11.928&0.282& 0.150&0.520&2.709& 1&  -   &  -   &  -   &  -   &032& 021&-012&-039&-031\\
96&  9.703&0.063& 0.219&0.976&2.887& 1&-     &  -   &  -   &   -  &037& 023&-043&-003&003\\
97& 11.782&0.129& 0.215&0.934&2.838& 1&-     &  -   &  -   &  -   &038& 015&-017&-034&017\\
102& 10.724&0.150& 0.193&0.893&2.851& 1&-     & -    &  -   &    - &036& 001& 001&001&-003\\
105& 11.166&0.167& 0.232&0.971&2.832& 1&-     & -    & -    &   -  &054& 009& 028&-175&-015\\
109& 10.034&0.040& 0.178&1.018&2.960& 1&-     &-     & -    & -    &-004& 026&-026&-005&-044\\
111&  9.900&0.025& 0.249&0.964&2.927& 1& -    & -    &  -   &  -   &050& 030&-086&057&-019\\
\enddata
\end{deluxetable}

\clearpage

\begin{deluxetable}{rrrrr}
\tablewidth{0pt}
\tablecaption{Mean catalog minus transformed values for the standard
stars and their standard deviations in $V$, (b$-$y), m$_1$, c$_1$ and $\beta$.}
\tablehead{
\colhead{$\Delta$$_m$(V)}&\colhead{$\Delta$$_m$(b$-$y)}&
\colhead{$\Delta$$_m$(m$_1$)}&\colhead{$\Delta$$_m$(c$_1$)}&\colhead{$\Delta$$_m$($\beta$)}}
\startdata
$0.014$&$0.003$&$-0.005$&$-0.004$&$-0.003$\\
$0.033$&$0.027$&$0.049$&$0.044$&$0.031$\\
\enddata
\end{deluxetable}

\clearpage

\begin{deluxetable}{lcc}
\tablewidth{0pt}
\tablecaption{Observational details of the optical spectroscopy. Observations 
were taken using the  1.52-m G.~D.~Cassini telescope and BFOSC. All dates
refer to February 2000.\label{tb2}}
\tablehead{
\colhead{Star} &\colhead{ Observation Date(s)} &\colhead{ Resolution}}
\startdata
S1R2N23 & 3/2 &  $220\:$\AA/mm\\
S1R2N35 & 3/2 &   $220\:$\AA/mm,  $170\:$\AA/mm\\
S2R1N16 & 3/2 &   $220\:$\AA/mm\\
S2R1N18 & 4/2 & $170\:$\AA/mm\\
S2R1N26 & 3/2, 6/2 & $220\:$\AA/mm\\
S2R2N43 & 3/2, 4/2 &  $220\:$\AA/mm,  $120\:$\AA/mm\\
S3R1N3 & 3/2 &  $220\:$\AA/mm\\
S3R1N16 & 4/2 & $170\:$\AA/mm\\
S3R1N17 & 4/2 & $170\:$\AA/mm\\
S3R2N15 & 4/2 & $170\:$\AA/mm\\
S4R2N14 & 3/2 &  $220\:$\AA/mm\\
S4R2N15 & 3/2 &  $220\:$\AA/mm\\
\enddata
\end{deluxetable}

\clearpage

\begin{deluxetable}{lcrrrcrrccc}
\tabletypesize{\small}
\tablewidth{0pt}
\tablecaption{Pixel coordinates and measured photometric parameters for all
the stars in the field of NGC~1893.\label{tb3}}
\tablehead{
\colhead{Star}           & \colhead{Frame} &
\colhead{X Position}     & \colhead{Y Position}&
\colhead{V}              & \colhead{b$-$y}&
\colhead{m$_1$}          & \colhead{c$_1$}&
\colhead{$\beta$}        & \colhead{N$_{uvby}$}&
\colhead{N$_\beta$}}
\startdata
S1R1N2  & \#2&745.8&783.4&  -  &   -  &  -   &  -  &2.511&-&1\\
S1R1N4  & \#2&799.5&818.9&15.707& 0.343& 0.127&0.977&2.948&2&2\\
S1R1N5  & \#2&831.8&854.8&  -  &   -  &  -   &  -  &2.560&-&1\\
S1R1N7  & \#3&678.5& 70.1&14.653& 0.207& 0.187&0.500&2.754&2&2\\
S1R1N8  & \#3&699.0&114.2&  -  &   -  &  -   &  -  &2.891&-&1\\
S1R1N9  & \#3&715.8&131.2&14.946& 0.540& 0.207&0.267&2.636&2&2\\
S1R1N10 & \#3&876.5& 43.9&14.604& 0.280& 0.033&0.583&2.745&1&1\\
S1R1N18 & \#2&856.6&694.4&14.469& 0.366& 0.223&0.413&2.633&1&1\\
S1R1N19 & \#2&760.6&642.0&11.100& 0.724& 0.607&0.249&2.587&1&1\\
S1R2N1  & \#2&820.3& 51.7&14.224& 0.278& 0.094&0.377&2.738&1&1\\
S1R2N2  & \#2&837.3& 60.2&13.744& 0.300& 0.228&0.728&2.752&1&1\\
S1R2N4  & \#2&977.5& 44.2&14.074& 0.421& 0.118&1.115&2.842&1&1\\
S1R2N6  & \#2&835.0&145.8&15.214& 0.313& 0.072&0.788&2.811&1&1\\
S1R2N19 & \#2&903.4&267.7&14.020& 0.418& 0.256&0.900&2.832&1&1\\
S1R2N20 & \#2&964.6&279.9&13.330& 0.356& 0.068&0.175&2.668&1&1\\
S1R2N21 & \#2&996.3&224.2&  -  &   -  &  -   &  -  &2.621&-&1\\
S1R2N22 & \#2&892.7&417.7&14.183& 0.361& 0.183&0.520&2.669&1&1\\
S1R2N23 & \#2&923.6&386.0&15.344& 0.655& 0.256&0.297&2.481&1&1\\
S1R2N25 & \#2&1013.&462.9&15.031& 0.398& 0.262&0.356&2.692&1&1\\
S2R1N1  & \#3&933.5&182.2&14.257& 0.272& 0.091&0.724&2.768&1&1\\
S2R1N2  & \#3&872.2&206.2&12.148& 0.199& 0.092&0.055&2.660&1&1\\
S2R1N3  & \#3&844.4&230.7&14.526& 0.267& 0.060&0.633&2.757&1&1\\
S2R1N4  & \#3&872.5&167.0&15.737& 0.589& 0.127&0.560&2.713&1&1\\
S2R1N7  & \#3&727.8&190.3&13.441& 0.182& 0.145&0.097&2.700&2&2\\
S2R1N9  & \#3&751.5&272.1&14.404& 0.342& 0.139&1.064&2.830&1&1\\
S2R1N11 & \#3&729.6&342.7&15.868& 0.467& 0.174&0.658&2.681&1&1\\
S2R1N12 & \#3&854.0&316.7&12.704& 0.234& 0.115&0.225&2.700&1&1\\
S2R1N15 & \#3&614.3&393.1&14.199& 0.193& 0.134&0.336&2.728&1&1\\
S2R1N16 & \#3&608.4&358.8&15.556& 0.793&-0.041&0.568&2.587&1&1\\
S2R1N18 & \#3&550.8&161.2&11.193& 0.161& 0.116&-0.054&2.644&2&2\\
S2R1N19 & \#3&518.7&179.4&15.007& 0.572& 0.119&0.616&2.705&2&2\\
S2R1N23 & \#3&540.8&374.2&13.705& 0.251& 0.055&0.362&2.685&1&1\\
S2R1N24 & \#3&556.2&400.3&12.722& 0.280& 0.037&0.274&2.707&1&1\\
S2R1N25 & \#3&466.4&480.8&13.048& 0.233& 0.049&0.123&2.663&1&1\\
S2R1N26 & \#3&503.4&502.0&15.386& 0.824& 0.066&0.408&2.555&1&1\\
S2R1N27 & \#3&557.0&514.4&15.059& 0.589& 0.252&0.391&2.592&2&2\\
S2R1N28 & \#3&556.6&565.8&15.083& 0.477& 0.121&0.308&2.644&2&2\\
S2R1N31 & \#3&517.1&626.0&14.439& 0.265& 0.153&1.024&2.862&2&2\\
S2R2N20 & \#3&900.0&670.5&13.171& 0.209& 0.114&0.173&2.713&1&1\\
S2R2N25 & \#3&935.2&725.7&15.403& 0.257& 0.177&0.832&2.889&1&1\\
S2R2N35 & \#3&742.2&1002.1&14.860& 0.383& 0.275&0.274&2.683&1&1\\
S2R2N40 & \#4&745.9&306.7&13.826& 0.322& 0.014&0.490&2.696&3&3\\
S2R2N41 & \#3&714.7&768.4&11.995& 0.163& 0.077&-0.030&2.656&3&3\\
S2R2N42 & \#3&672.0&825.7&12.383& 0.186& 0.092&0.128&2.680&3&3\\
S2R2N43 & \#3&659.1&854.0&11.465& 0.167& 0.068&-0.054&2.633&3&3\\
S2R2N45 & \#4&840.0&432.7&14.483& 0.329& 0.166&0.981&2.884&2&2\\
S2R2N46 & \#4&778.1&534.2&12.514& 0.525& 0.112&0.461&2.601&2&2\\
S2R3N59 & \#4&773.6&824.1&15.251& 0.396& 0.085&1.073&2.925&1&1\\
S3R1N1  & \#3&387.5&614.5&12.269& 0.258&-0.004&0.151&2.632&3&3\\
S3R1N2  & \#3&355.5&554.5&13.664& 0.246& 0.040&0.348&2.689&3&3\\
S3R1N3  & \#3&305.8&488.8&13.612& 0.603&-0.161&0.156&2.414&1&1\\
S3R1N4  & \#3&422.6&510.5&13.948& 0.284& 0.060&0.323&2.596&3&3\\
S3R1N6  & \#3&424.5&440.5&12.485& 0.293& 0.012&0.341&2.678&1&1\\
S3R1N8  & \#3&425.1&370.7&13.738& 0.284& 0.028&0.573&2.719&1&1\\
S3R1N9  & \#3&221.7&552.0&  -  &   -  &   -  &  -  &2.644&-&1\\
S3R1N12 & \#3&147.8&339.0&12.472& 0.373& 0.013&0.780&2.734&1&1\\
S3R1N13 & \#3&151.3&243.4&13.151& 0.425&-0.059&0.673&2.758&1&1\\
S3R1N16 & \#3&340.5&242.3&10.273& 0.257&-0.052&-0.060&2.598&1&1\\
S3R1N17 & \#3&369.5&211.5&11.339& 0.213& 0.077&-0.083&2.634&2&2\\
S3R2N1  & \#4&681.2&506.1&10.774& 0.148& 0.168&1.072&2.837&3&3\\
S3R2N2  & \#4&700.4&322.4&12.579& 0.234& 0.047&0.370&2.676&3&3\\
S3R2N4  & \#4&631.4&372.6&13.530& 0.282& 0.020&0.466&2.694&3&3\\
S3R2N5  & \#4&536.2&366.4&14.420& 0.395&-0.003&0.463&2.712&2&2\\
S3R2N6  & \#3&435.0&659.4&14.756& 0.604& 0.052&0.894&2.705&2&2\\
S3R2N7  & \#3&443.7&646.0&14.336& 0.428& 0.026&1.123&3.027&2&2\\
S3R2N8  & \#4&484.3&443.6&15.049& 0.490&-0.037&0.566&3.348&2&2\\
S3R2N9  & \#4&516.0&446.0&15.245& 0.493&-0.033&0.637&2.756&2&2\\
S3R2N13 & \#4&420.1&616.5&15.286& 0.450&-0.095&0.837&2.777&1&1\\
S3R2N15 & \#4&469.5&506.1&10.161& 0.420&-0.073&-0.023&2.592&3&3\\
S3R2N16 & \#4&338.5&509.6&14.738& 0.624&-0.001&1.279&2.927&2&2\\
S3R2N17 & \#4&374.1&468.6&15.063& 0.481& 0.153&0.404&2.664&2&2\\
S3R2N18 & \#4&364.2&341.9&12.907& 0.404&-0.071&0.152&2.646&3&3\\
S3R2N26 & \#3& 66.2&570.1&13.456& 0.321& 0.099&0.965&2.801&3&3\\
S3R2N27 & \#4&282.9& 50.8&14.204& 0.371& 0.065&0.526&2.648&1&1\\
S3R3N7  & \#4&505.9&867.2&13.293& 0.419& 0.154&0.559&2.695&2&2\\
S3R3N11 & \#4&532.4&682.9&14.502& 0.474&-0.070&0.438&2.671&1&1\\
S3R3N12 & \#4&242.0&726.8&14.383& 0.551& 0.050&0.376&2.600&1&1\\
S3R3N13 & \#4&211.8&832.9&15.177& 0.540&-0.119&0.625&2.761&1&1\\
S4R1N1  & \#2&555.1&866.2&14.712& 0.519& 0.173&0.312&2.629&2&2\\
S4R1N4  & \#2&478.1&772.8&13.956& 0.345& 0.231&0.450&2.677&1&1\\
S4R1N8  & \#2&461.5&649.8&13.184& 0.386& 0.163&0.422&2.674&3&3\\
S4R1N10 & \#2&559.0&608.7&14.905& 0.265& 0.109&0.637&2.841&2&2\\
S4R1N11 & \#2&588.8&608.5&14.793& 0.398& 0.230&0.346&2.650&1&1\\
S4R1N13 & \#2&714.7&613.4&14.388& 0.420& 0.243&0.387&2.617&1&1\\
S4R1N16 & \#2&337.2&595.2&15.519& 0.285& 0.141&0.936&2.862&2&2\\
S4R1N17 & \#2&480.9&478.2&11.780& 0.225& 0.233&0.794&2.824&3&3\\
S4R2N3  & \#1&226.5&950.8&14.475& 0.249& 0.262&0.820&2.930&1&1\\
S4R2N7  & \#1&373.9&951.7&14.239& 0.424& 0.111&1.226&2.888&1&1\\
S4R2N9  & \#2& 97.7&747.9&14.985& 0.488& 0.200&0.375&2.676&2&2\\
S4R2N12 & \#2&106.4&494.5&15.333& 0.504& 0.032&0.550&2.702&2&2\\
S4R2N14 & \#1&430.3&572.3&15.482& 0.518& 0.123&0.374&2.575&1&1\\
S4R2N15 & \#1&381.2&549.0&14.488& 0.592& 0.447&0.320&2.514&1&1\\
S4R2N17 & \#1&500.0&500.0&9.036& 0.266&-0.022&-0.123&2.579&1&1\\
S4R2N18 & \#2&247.7&506.7&12.297& 0.253& 0.028&0.230&2.664&3&3\\
S4R2N19 & \#2&337.0&452.5&12.284& 0.161& 0.204&1.026&2.895&3&3\\
S4R2N20 & \#2&403.7&424.0&14.606& 0.448& 0.188&0.384&2.627&2&2\\
S4R2N22 & \#2&579.6&340.6&13.856& 0.342& 0.192&0.652&2.707&3&3\\
S4R2N23 & \#2&289.3&333.1&15.613& 0.488& 0.065&0.468&2.637&2&2\\
S4R2N24 & \#2&466.2&256.1&15.820& 0.424& 0.213&0.220&2.627&1&1\\
S4R2N27 & \#2&659.7&183.4&13.702& 0.401& 0.208&0.351&2.640&1&1\\
S4R2N28 & \#2&716.8&204.1&13.286& 1.022& 0.583&0.372&2.574&1&1\\
S4R2N29 & \#1&598.3&380.9&14.490& 0.591& 0.408&0.401&2.571&2&2\\
S4R2N30 & \#1&647.2&334.4&13.645& 0.291& 0.166&0.849&2.746&3&3\\
S4R2N32 & \#1&826.7&364.7&14.255& 0.411& 0.168&0.464&2.655&2&2\\
S4R2N33 & \#1&783.0&286.6&14.646& 0.291& 0.181&0.976&2.842&1&1\\
S4R2N34 & \#1&818.0&285.3&14.037& 0.386& 0.182&0.461&2.666&2&2\\
S4R3N2  & \#1& 81.6&971.3&11.728& 0.192& 0.076&0.229&2.690&2&2\\
S4R3N20 & \#1&323.2&466.0&13.174& 0.555& 0.320&0.363&2.588&2&2\\
S4R3N23 & \#1&171.4&158.0&14.544& 0.844& 0.177&0.362&2.580&1&1\\
S4R3N24 & \#1&134.2&108.6&14.695& 0.509& 0.191&0.374&2.564&1&1\\
S4R3N29 & \#1&602.6&138.9&12.782& 0.493& 0.140&0.434&2.615&2&2\\
S4R3N32 & \#1&654.2&297.3&  -  &   -  &   -  &  -  &2.690&-&1\\
S5000   & \#3&426.2&361.6&14.149& 0.270& 0.036&0.499&2.721&1&1\\
S5001   & \#3&859.1&309.1&15.116& 0.565& 0.157&0.274&2.557&1&1\\
\enddata
\end{deluxetable}

\clearpage

\begin{deluxetable}{lrrrccc}
\tablewidth{0pt}
\tablecaption{Unreddened indices and an estimation of the spectral classification for all the likely members of NGC~1893. \label{tb4}}
\tablehead{
\colhead{Star}&\colhead{[c$_1$]}&\colhead{[m$_1$]}&\colhead{[u$-$b]}&\colhead{T$_{\rm eff}$}&\colhead{$Spectral Type$}&\colhead{$Spectroscopic$}
}
\startdata
\cutinhead{O$-$Type}
S3R1N16&$-0.111$&$0.030$&$-0.051$&$-$&$-$&O8V(f)$^{1,2}$, O7V((f))$^{3,5}$\\
S4R2N17&$-0.176$&$0.063$&$-0.050$&$-$&$-$&O6V(f)$^{1}$, O5$^{4}$\\
S3R2N15&$-0.107$&$0.061$&$0.016$&$-$&$-$&O8$^2$, O6$^{3}$, O5V(f)$^{5}$\\
\cutinhead{B$-$Type}
S2R2N43&$-0.087$&0.121&0.155&23710&B1&B1III$^{1}$, B0.5V$^{5}$\\
S3R1N17&$-0.126$&0.145&0.164&23436&B1&B0.5V$^{1}$, B0.2V$^{5}$\\ 
S3R2N18& 0.071& 0.058& 0.188&22736&B1.5&B1.5V$^{1}$\\ 
S2R2N41&$-0.063$&0.129&0.196&22512&B1.5&B1.5V$^{1,5}$\\ 
S2R1N18&$-0.086$&0.168&0.249&21140&B2&B1V$^{1}$, B0.7V$^{5}$\\
S3R1N1&0.099&0.079&0.257&20948&B2&B2.5V$^{1}$, B2V$^{5}$\\ 
S2R1N25&0.076&0.124&0.324&19476&B2&\\  
S2R1N2&0.015&0.156&0.327&19415&B2& B1V$^{5}$\\  
S2R2N42&0.091&0.152&0.394&18156&B3&\\
S4R2N18&0.179&0.109&0.397&18104&B3&\\ 
S4R3N2&0.191&0.137&0.465&16999&B4&B2V$^{3}$\\
S2R1N7&0.061&0.203&0.467&16969&B4&\\ 
S1R2N20&0.104&0.182&0.468&16954&B4&\\
S2R1N24&0.218&0.127&0.471&16909&B4&\\ 
S2R2N20&0.131&0.181&0.493&16585&B4&\\
S3R1N6&0.282&0.106&0.494&16571&B4&\\ 
S3R3N11&0.343&0.082&0.507&16387&B4&\\
S3R1N2&0.299&0.119&0.536&15991&B5&\\
S2R1N12&0.178&0.190&0.558&15704&B5&\\ 
S3R2N2&0.323&0.122&0.567&15590&B5&\\ 
S2R1N23&0.312&0.135&0.582&15404&B5&\\
S3R2N4&0.410&0.110&0.630&14840&B6&\\  
S3R2N5&0.384&0.123&0.631&14829&B6&\\
S2R2N40&0.426&0.117&0.660&14510&B6&\\ 
S1R2N1&0.321&0.183&0.687&14226&B6&\\ 
S2R1N15&0.297&0.196&0.689&14206&B6&\\ 
S5000&0.445&0.122&0.690&14195&B7&\\ 
S3R1N13&0.588&0.077&0.742&13684&B8&\\ 
S3R1N8&0.516&0.119&0.754&13572&B8&\\ 
S1R1N10&0.527&0.123&0.772&13408&B8&\\
S3R2N9&0.538&0.125&0.788&13265&B8&\\ 
S3R2N13&0.747&0.049&0.845&13784&B9&\\
S2R1N3&0.580&0.145&0.870&12585&B9&\\
S1R1N7&0.459&0.253&0.965&11888&B9&\\ 
S4R1N10&0.584&0.194&0.972&11840&B9&\\
S2R1N1&0.670&0.178&1.026&11485&B9&\\
S1R2N6&0.725&0.172&1.070&11213&B9&\\
\cutinhead{A$-$Type}
S2R2N25&0.781&0.259&1.299&10007&A0&\\ 
S4R1N16&0.879&0.232&1.343&9809&A0&\\  
S1R1N4&0.908&0.237&1.382&9641&A1&\\   
S2R3N59&0.994&0.212&1.417&9496&A1&\\  
\tablerefs{
$^1$Massey et al. (1995); $^2$Hoag et al. (1965); $^3$Hiltner (1966); $^4$Hiltner (1956); $^5$This work
}
\enddata
\end{deluxetable}

\clearpage

\begin{deluxetable}{lrrrrrc}
\tablewidth{0pt}
\tablecaption{Intrinsic photometric parameters for all the likely members of 
NGC~1893. The color excess $E(b-y)$ has been calculated by applying
Crawford's (1978) procedure. The average cluster reddening has been used
to derive the intrinsic photometric parameters. The absolute magnitude
and distance modulus are calculated by using the calibration of
Balona \& Shobbrook (1984), based on the value of the $\beta$ index.\label{tb5}}
\tablehead{
\colhead{Star}&\colhead{E(b$-$y)}&\colhead{V$_0$}&\colhead{(b$-$y)$_0$}&\colhead{c$_0$}&\colhead{M$_V$($\beta$)}&\colhead{V$_0$$-$M$_V$($\beta$)}
}
\startdata
S1R1N4  &0.371& 14.287&  0.013& 0.911& 1.674&     12.613\\
S1R1N7  &0.280& 13.233& -0.123& 0.434& -0.216&    13.449\\
S1R1N10 &0.346& 13.184& -0.050& 0.517& -0.305&    13.489\\
S1R2N1  &0.365& 12.804& -0.052& 0.311& -0.529&    13.333\\
S1R2N6  &0.360& 13.794& -0.017& 0.722& 0.475&     13.319\\
S1R2N20 &0.464& 11.911&  0.026& 0.109&-1.874&     13.784\\
S2R1N1  &0.324& 12.837& -0.058& 0.658& -0.030&    12.867\\
S2R1N2  &0.316& 10.728& -0.131&-0.011& -2.478&    13.206\\
S2R1N3  &0.328& 13.106& -0.063& 0.567& -0.148&    13.254\\
S2R1N7  &0.294& 12.021& -0.148& 0.031& -1.757&    13.778\\
S2R1N12 &0.335& 11.284& -0.096& 0.159& -1.292&    12.576\\
S2R1N15 &0.282& 12.779& -0.137& 0.270& -0.711&    13.490\\
S2R1N18 &0.288&  9.773& -0.169&-0.120& -3.389&    13.162\\
S2R1N23 &0.338& 12.285& -0.079& 0.296& -1.276&    13.561\\
S2R1N24 &0.377& 11.302& -0.050& 0.208& -1.085&    12.387\\
S2R1N25 &0.344& 11.628& -0.097& 0.057& -2.127&    13.755\\
S2R2N20 &0.314& 11.751& -0.121& 0.107& -1.304&    13.055\\
S2R2N25 &0.298& 13.983& -0.073& 0.766& 1.213&     12.770\\
S2R2N40 &0.398& 12.406& -0.008& 0.424& -1.057&    13.463\\
S2R2N41 &0.287& 10.575& -0.167&-0.096& -3.034&    13.609\\
S2R2N42 &0.295& 10.963& -0.144& 0.062& -1.869&    12.832\\
S2R2N43 &0.294& 10.045& -0.163&-0.120& -3.593&    13.638\\
S2R3N59 &0.410& 13.831&  0.066& 1.007& 1.479&     12.352\\
S3R1N1  &0.366& 10.849& -0.072& 0.085& -2.585&    13.434\\
S3R1N2  &0.335& 12.244& -0.084& 0.282& -1.223&    13.467\\
S3R1N4  &0.376& 12.528& -0.046& 0.257& -3.771&    16.299\\
S3R1N6  &0.383& 11.065& -0.037& 0.275& -1.412&    12.477\\
S3R1N8  &0.351& 12.318& -0.046& 0.507& -0.690&    13.008\\
S3R1N16 &0.386& 8.854 & -0.073&-0.126& -4.405&    13.258\\
S3R1N17 &0.344& 9.919& -0.117&-0.149& -3.767&    13.686\\
S3R2N2  &0.320& 11.159& -0.096& 0.304& -1.429&    12.588\\
S3R2N4  &0.360& 12.110& -0.048& 0.400& -1.089&    13.199\\
S3R2N5  &0.475& 13.000&  0.065& 0.397& -0.795&    13.795\\
S3R2N9  &0.558& 13.825&  0.163& 0.571& -0.162&    13.987\\
S3R2N13 &0.494& 13.866&  0.120& 0.771& 0.019&     13.847\\
S3R2N15 &0.549&  8.742&  0.090&-0.089& -4.289&    13.030\\
S3R2N18 &0.515& 11.487&  0.074& 0.086& -2.308&    13.795\\
S3R3N11 &0.558& 13.082&  0.144& 0.372& -1.526&    14.608\\
S3R3N13 &0.607& 13.757&  0.210& 0.559& -0.095&    13.852\\
S4R1N10 &0.326& 13.485& -0.065& 0.571& 0.740&     12.745\\
S4R1N16 &0.316& 14.099& -0.045& 0.870& 0.959&     13.140\\
S4R2N17 &0.402& 7.617 & -0.064&-0.189& -5.555&    13.171\\
S4R2N18 &0.354& 10.877& -0.077& 0.164& -1.802&    12.679\\
S4R3N2  &0.291& 10.308& -0.138& 0.163& -1.411&    11.719\\
S5000   &0.344& 12.729& -0.060& 0.433& -0.652&    13.381\\
\enddata 
\end{deluxetable}

\clearpage

\begin{deluxetable}{lccc}
\tablewidth{0pt}
\tablecaption{Photometric intrinsic indices for stars which are likely to be
pre-main-sequence members of NGC~1893.\label{tb6}}
\tablehead{
\colhead{Star}&\colhead{V$_0$}&\colhead{(b$-$y)$_0$}&\colhead{c$_0$}
}
\startdata
S1R2N23 & 13.924&  0.325& 0.231\\
S2R1N16 & 14.136&  0.463& 0.502\\
S2R1N26 & 13.966&  0.494& 0.342\\
S3R1N3  & 12.192&  0.273& 0.090\\
S4R2N14 & 14.062&  0.188& 0.308\\
S4R2N15 & 13.068&  0.262& 0.254\\
S5001   & 13.696&  0.235& 0.208\\
\enddata
\end{deluxetable}

\end{document}